\documentclass[12pt]{elsarticle}
	\usepackage{natbib}
	\usepackage{changes}
	\usepackage{booktabs} 
	\usepackage{graphicx} 
	\usepackage{amsmath} 
	\usepackage{enumerate}
	\usepackage{epsfig}
	\usepackage[export]{adjustbox}
	\usepackage{amssymb}
	\usepackage[utf8]{inputenc}
	\usepackage{booktabs}
	\usepackage{listings}
	\usepackage{color, colortbl}
	\usepackage{graphicx}
	\usepackage{svg}
	\usepackage{epstopdf}
	\usepackage{caption}
	\usepackage{amsmath,empheq,amssymb,mathtools}
	\usepackage{gensymb}
	\usepackage{float}
	\usepackage{subfloat}
	\usepackage{subfig}
	\usepackage{mathtools}
	\usepackage{cancel}
	\usepackage{moreverb,url}
	\usepackage{enumitem}
	\usepackage{physics}
	\usepackage{xcolor}
	
	\newtheorem{thm}{Theorem}
	
\begin{document}
	
\begin{frontmatter}		
\title{Resetting Disturbance Observers with application in Compensation of bounded nonlinearities like Hysteresis in Piezo-Actuators \footnote{Corresponding author: S. Hassan HosseinNia, Department of Precision and Microsystems Engineering, Delft University of Technology, Mekelweg 2, 2628 CD Delft, The Netherlands. Email: \url{s.h.hosseinniakani@tudelft.nl}}} 

\author[label1]{Niranjan Saikumar}
\author[label1]{Rahul Kumar Sinha}
\author[label1]{and S. Hassan HosseinNia}
\address[label1]{Department of Precision and Micro System Engineering, Delft University of Technology, The Netherlands}

\begin{abstract}
{

{
This paper presents a novel nonlinear (reset) disturbance observer for dynamic compensation of bounded nonlinearities like hysteresis in piezoelectric actuators. Proposed Resetting Disturbance Observer (RDOB) utilizes a novel Constant-gain Lead-phase (CgLp) element based on the concept of reset control. The fundamental limitations of linear DOB which results in contradictory requirements and in a dependent design between DOB and feedback controller are analysed. Two different configurations of RDOB which attempt to alleviate these problems from different perspectives are presented and an example plant is used to highlight the improvement. Stability criteria are presented for both configurations. Performance improvement seen with both RDOB configurations compared to linear DOB is also verified on a practical piezoelectric setup for hysteresis compensation and results analysed.
}
}
\end{abstract}


\end{frontmatter}

\newpage

\section{Introduction}

Real world is nonlinear and plants are often represented by their nominal linear model. All existing nonlinearities are generally considered as input disturbances and are handled by the disturbance rejection capability of the overall feedback system. Linear controllers are designed based on the nominal linear plant estimate and in the absence of a disturbance observer (DOB), the effect of nonlinearities have to be dealt solely by the disturbance rejection property of the controller. In some cases where the nonlinearities are significantly high to be handled by the controller, overall system in best case gets inaccurate and, in worst case can be unstable. One such nonlinear system is a piezo-actuator where high nonlinearity exists due to its hysteresis. Piezo-actuators have become increasingly popular in high precision motion control applications \cite{woronko2003piezoelectric,gu2014high}, and literature focussing on hysteresis compensation specifically in piezo-actuators through DOB can be found in \cite{gu2014high,sofla2010hysteresis,abidi2004sliding,yi2009disturbance,ruderman2014control}. In general, suppression of similar nonlinear effects has been widely studied using various techniques in literature and can be broadly classified into two categories namely, model-based techniques and control based techniques.

Models used in the model-based techniques can be either physics principles-based, differential equation based, mathematical operator based, fuzzy logic based and so on \cite{jiles1986theory,al2009hysteresis,al2010compensation,gu2016modeling,hassani2014survey,janaideh2009generalized,krejci2001inverse,rakotondrabe2012classical}. They operate on a similar mode by modelling the nonlinearity and using the inverse which is connected in series to have a feedforward connection such that the nonlinearity gets cancelled. The main advantage of this approach is that it is a feedforward compensation scheme and can hence not cause instabilities. However, an equally major drawback is that it is not generalizable since a new model has to be identified for each device and worse for every new operating condition and this significantly limits its utilization in industrial applications. Additionally, the achievable accuracy is determined by the accuracy of nonlinearity estimate. These techniques are not analysed in this work due to their lack of robustness.

On the other hand, control based techniques can be classified into two categories namely, feedforward and feedback. Feedforward control strategy is same as model-based technique. However, the feedback approach does not need the hysteresis model and is hence generalizable resulting in enhanced compensation of nonlinearity with a single compensation architecture operating at all different operating points. This strategy generally considers nonlinearities to be bounded input disturbances and attempts to reject them. This methodology can be further classified into two categories, one which estimates considered nonlinearity as disturbance using linear state estimation \cite{yi2009disturbance} and, another which uses nominal linear plant model and attenuates all other dynamics including actual input disturbance \cite{du2010simple}. It is not clear as to why nonlinearity is estimated as nonlinear disturbance using linear estimation techniques in the first approach since this limits its performance. However, the performance of the second approach is more limited by fundamental properties of linear control systems and not the approach itself and this second approach is the focus of our work. 

{
Linear disturbance observers (DOB) suffer from the fundamental limitations. One of the limitations comes from the contradictory requirements for disturbance rejection and noise attenuation placed on the disturbance estimating filter (DEF) which is an integral part of DOB \cite{schrijver2002disturbance}. Another is that, although ideally design of feedback controller and DOB should be separable, the limitation on sensitivity function results in a dependence as explained in detail in the next section. This paper presents novel nonlinear DOB configurations to specifically tackle these limitations and the nonlinearity used for this purpose is reset control. 
}

{
Reset control is a nonlinear control technique which has been the focus of research for many decades starting from Clegg in 1958 \cite{clegg1958nonlinear}. The phase advantage of reset has been used to overcome limitations of linear control in recent years \cite{banos2011reset,zheng2008development}. Its simplicity along with the fact that it can be approximated and analysed in frequency domain using describing function gives it a great advantage over other nonlinear techniques. Several works exist where reset has been used for performance improvement in high precision motion control \cite{li2011optimal,li2011reset}. However, application of reset in DOB for performance improvement does not exist to the best of authors' knowledge. In this paper, reset control is applied for the first time in DOB to improve performance. A novel reset element 'Constant-gain Lead-phase' is used for this purpose and two different approaches are considered to improve performance resulting in two different novel configurations being presented.
}

{
Hysteresis compensation for piezo-actuators is considered for application of proposed schemes. It must be noted that the work focusses on overcoming fundamental limitations of linear DOB through the introduction of reset. Although hysteresis compensation has been chosen as the application example, advancement on this front is not considered. As such, comparison of proposed scheme with advanced compensation schemes tailored for hysteresis compensation of piezo-actuators is not considered. The proposed schemes are only compared with linear DOB to show that the identified fundamental limitations are overcome.
}

The structure of the paper is as follows. The basics of nonlinear plant with bounded input disturbance and stability criteria of the feedback disturbance observer scheme proposed in \cite{yi2009disturbance} are studied in Section. \ref{NPDOB}, along with limitations of linear approach. The preliminaries of reset control along with design concept of `Constant-gain Lead-phase' (CgLp) element are explained in Section. \ref{Resetprel}. In Section. \ref{RDOB}, CgLp is used to present two novel Resetting Disturbance Observer (RDOB) configurations. A simple example plant is used to show that these overcome limitations of linear DOB. In Section. \ref{PZ}, hysteresis compensation in piezo-actuator is considered and results from the experimental set-up are provided for validation of proposed RDOB.  

\section{Nonlinear plant model and linear disturbance observers}
\label{NPDOB}

\begin{figure}[!htpb]
	\centering
	\includegraphics[width=\linewidth, keepaspectratio]{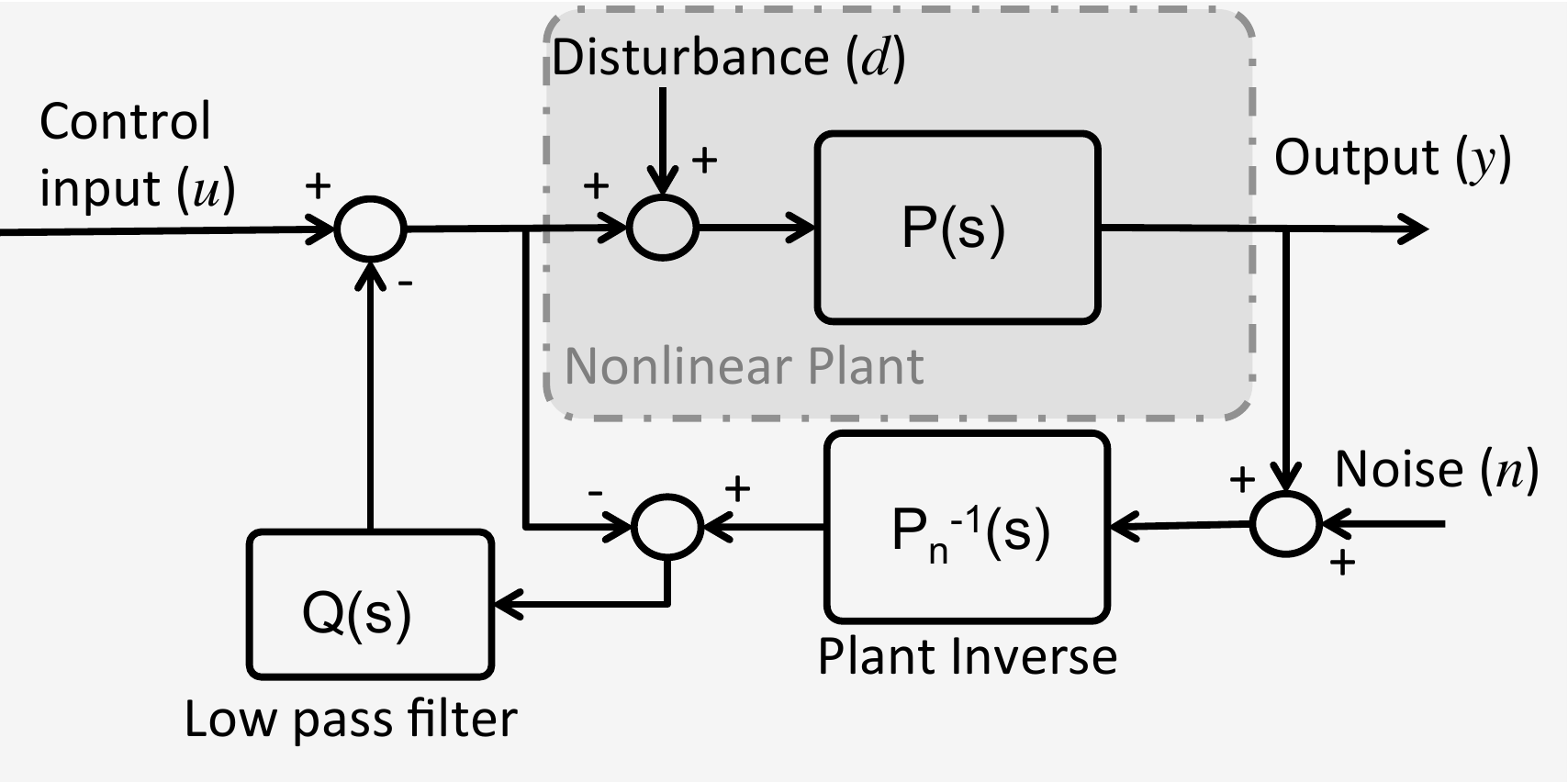}
	\caption{Disturbance observer architecture}
	\label{fig:DOB}	
\end{figure}

A plant having nonlinearities such as hysteresis, creep, or electromagnetic effects can be modelled as a combination of linear plant $P$ and bounded disturbance $d$ depicting the nonlinearities as shown in Fig. \ref{fig:DOB} \cite{schrijver2002disturbance}. The boundedness of disturbance is an important criterion and for instance, hysteretic nonlinearities can be proven to be bounded using Duhem model \cite{yi2009disturbance}. Similarly, other mentioned nonlinear effects can also be proven to follow the bounded input bounded output (BIBO) property and this model holds for any other BIBO nonlinear effects.

Disturbance observer has been proposed in \cite{shahruz2000performance} to compensate for large nonlinear effects. This is also capable of suppressing other input/output disturbances and unmodelled dynamics. A good DOB is capable of ensuring that the plant behaves like the nominal linear plant model and this linearization further allows for a relaxation on the robustness constraint of the feedback controller and also enhances the overall performance. The traditional compensation architecture of the disturbance observer as suggested in \cite{schrijver2002disturbance} is shown in Fig. \ref{fig:DOB}. Here, $P_n$ is the nominal linear model estimate of the plant and $Q$ is the disturbance estimating filter (DEF). 

\subsection{General Required Behaviour of $Q$-filter}
The transfer functions from the various inputs $u,n,d$ to output $y$ of the compensated scheme are derived in \cite{schrijver2002disturbance} as

\begin{align}
H_{uy}=\frac{PP_n}{Q(P-P_n)+P_n}
\label{tf1}\\
H_{ny}=\frac{PQ}{Q(P-P_n)+P_n}
\label{sens1}\\
H_{dy}=\frac{PP_n(1-Q)}{Q(P-P_n)+P_n}
\label{dis1}
\end{align}

Assuming $P = P_n$

\begin{gather}
H_{uy}=P_n
\label{tf2}\\
H_{ny}=Q
\label{sens2}\\
H_{dy}=P_n(1-Q)
\label{dis2}
\end{gather}

\begin{figure}[!htpb]
	\centering
	\includegraphics[width=\linewidth, keepaspectratio]{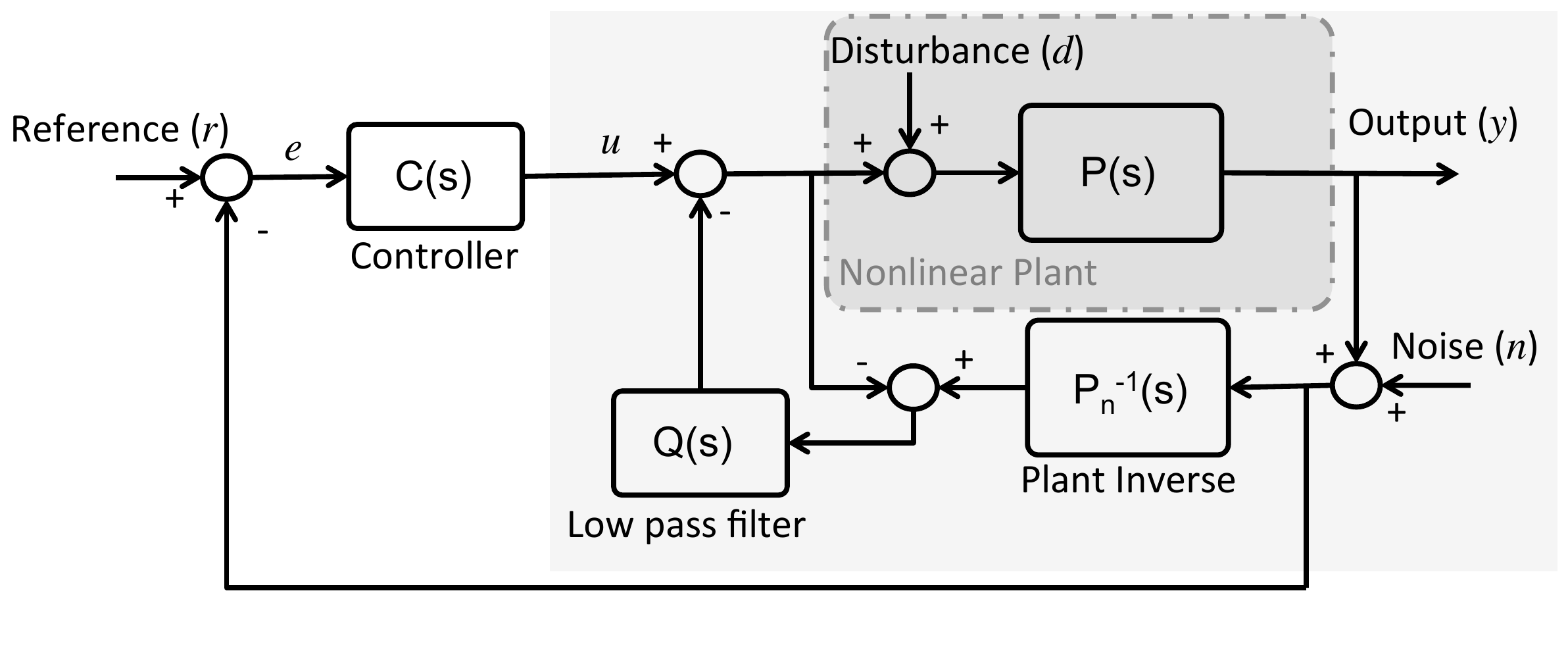}
	\caption{Full DOB with controller}
	\label{fig:DOB+cont}	
\end{figure}

In the case of a perfect match as above, $H_{uy}$ is equal to $P_n$ (nominal plant) and hence the feedback controller that is used in conjunction with DOB (see Fig. \ref{fig:DOB+cont}) sees only $P_n$. In this ideal scenario, DOB and feedback controller $C$ can be designed independently (also known as separation property).

Some required properties of DEF $Q$ can also be ascertained from the above equations. From Eqn. \ref{dis2}, complete disturbance rejection requires $Q$ to be as close as possible to unity. However, from Eqn. \ref{sens2}, complete noise attenuation requires $Q$ to be as close as possible to zero. These two requirements contradict each other and thereby limit the performance of compensation scheme. In the case under consideration, we are interested in countering hysteresis which can be considered as low frequency disturbance and we are also interested in high frequency noise attenuation. Hence, we can conclude that $Q$ should be a low pass filter. Additionally, to enable practical implementation, $Q$ should have a relative order equal to or higher than the nominal linear plant model $P_n$. All these requirements taken together significantly restrict design of $Q$ and thereby compensation to a certain frequency range.

\subsection{Inseparability of controller $C$ and DOB design}

\begin{figure}[!htpb]
	\centering
	\includegraphics[width=\linewidth, keepaspectratio]{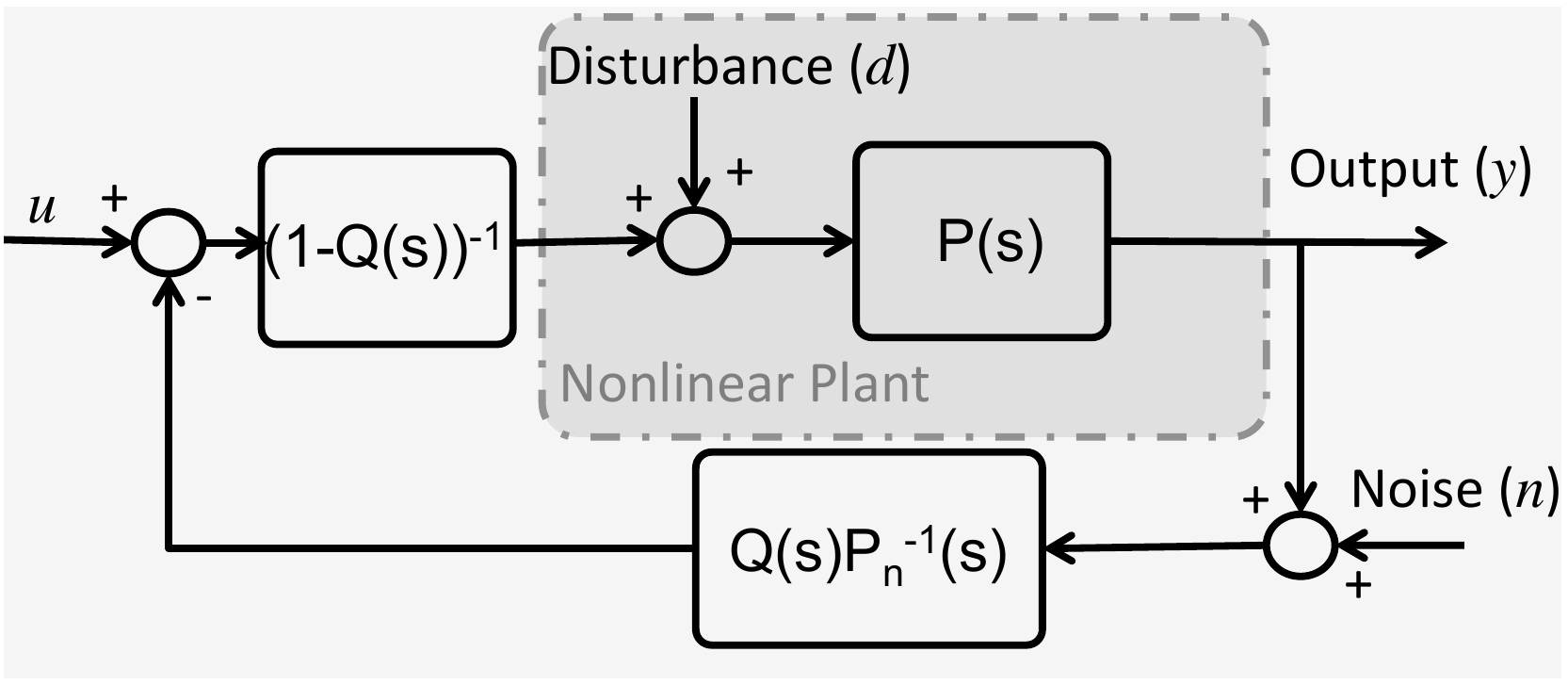}
	\caption{An equivalent representation of DOB}
	\label{fig:equivDOB}	
\end{figure}

{
Although in the ideal case, controller $C$ sees only nominal plant $P_n$ leading to separation principle, a more detailed sensitivity analysis shows that this is not the case. For this, the DOB scheme in Fig. \ref{fig:DOB} can equivalently be represented as shown in Fig. \ref{fig:equivDOB} for convenience according to \cite{schrijver2002disturbance}. From this figure, we can obtain sensitivity $S = 1/(1 + L)$ and complementary sensitivity $T = L/(1 + L)$ where $L$ is loop gain as
}

\begin{gather}
T = \frac{QP}{Q(P - P_n) + P_n}\\
S = \frac{P_n(1 - Q)}{Q(P - P_n) + P_n}
\end{gather}

{
Again assuming $P = P_n$, we get $T = Q$ and $S = (1 - Q)$ highlighting the importance of $Q$ design in performance of DOB.
}

{
Now if we consider the complete picture with controller $C$ as in Fig. \ref{fig:DOB+cont}, the realized transfer functions are given as explained in \cite{schrijver2002disturbance} as (assuming $P = P_n$):
}

\begin{gather}
H_{ry} = \frac{P_nC}{1 + P_nC} = T_c
\label{cleq1}\\
H_{dy} = \frac{P_n(1 - Q)}{1 + P_nC} = PS_c(1 - Q) = PS
\label{cleq2}\\
H_{ny} = \frac{P_nC + Q}{1 + P_nC} = T_c + \frac{Q}{1 + P_nC}
\label{cleq3}
\end{gather}
where 

\begin{gather}
S_c = 1/(1 + P_nC)
\label{eq_sens_outer}\\
T_c = P_nC/(1 + P_nC)
\end{gather}

{
are sensitivity and complementary sensitivity functions of controlled system without DOB respectively.
}

While from Eqns. \ref{tf1} to \ref{dis2} it is clear that $Q$ design is critical in DOB performance, from Eqns. \ref{cleq1} to \ref{cleq3} it can be seen that $Q$ is critical for performance of the overall closed-loop system as well. Further from Eqns. \ref{cleq1} and \ref{cleq2}, it is clear that peaking of $1 - Q$ and $Q$ should not coincide with peaking of $S_c$ and $T_c$ (which are realized by controller without DEF) respectively to ensure that peaking of overall sensitivity is limited. Hence, although ideally, separation principle holds, DEF and controller $C$ cannot be designed independently.

This also further restricts design of $Q$ affecting disturbance rejection performance. Ideally, $Q$ with corner frequency close to bandwidth of closed-loop system would be preferred to reject disturbances till bandwidth and attenuate noise at higher frequencies. However, such an approach would result in overlap of sensitivity peaks. This limitation is highlighted through an introductory example.

Consider a plant whose nominal linear model is given as a second order low pass filter
\begin{equation}
P_n = \frac{1}{{(s/\omega_p)}^2 + (2 s/\omega_p) + 1}
\label{eq_pn}
\end{equation}
A series PID controller can be designed for this system to obtain a bandwidth of $\omega_o$ as
\begin{equation}
C = K\Bigg(\frac{s/\omega_i + 1}{s}\Bigg)\Bigg(\frac{s/\omega_1 + 1}{s/\omega_2 + 1}\Bigg)\Bigg(\frac{1}{s/\omega_f + 1}\Bigg)
\label{eq_c}
\end{equation}
where $\omega_i < \omega_1 < \omega_o < \omega_2 < \omega_f$ and K is the dc gain. All values are chosen to obtain the required phase and gain margin.

Since the order of $Q$ has to be higher or equal to that of $P_n$, let $Q$ be a damped second order LPF of form
\begin{equation}Q = \frac{1}{{(s/\omega_q)}^2 + 2s/\omega_q + 1}
\label{eq_q}
\end{equation}

\begin{figure}[!htpb]
	\centering
	\includegraphics[width=\linewidth, keepaspectratio]{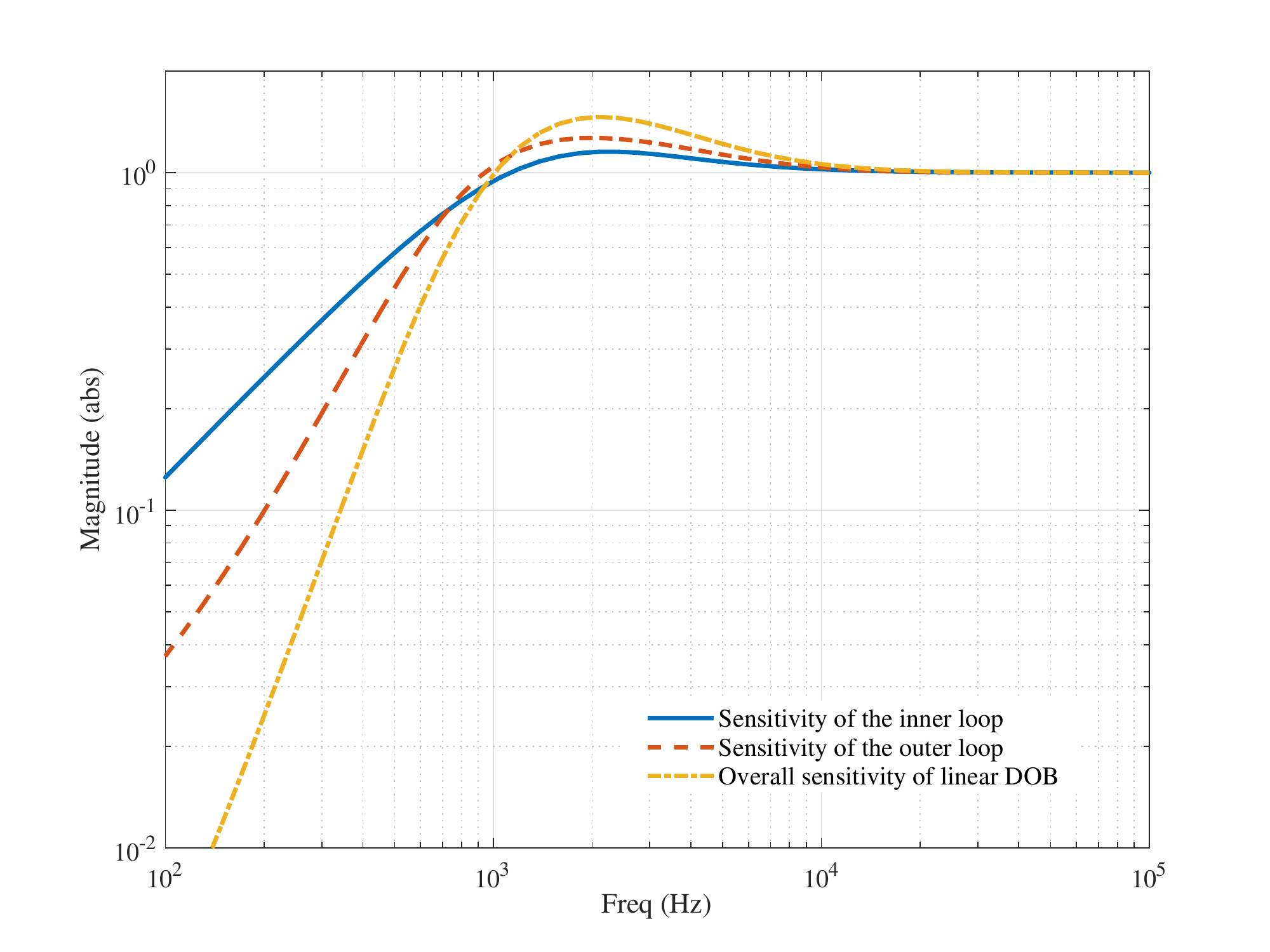}
	\caption{Sensitivity function plots with linear DOB}
	\label{fig:Example_1}	
\end{figure}

To obtain a sense of this in frequency domain, let us consider $\omega_p = 1000$ and to obtain a bandwidth of around $1.59 KHz = 1e4\ rad/s$, we get $\omega_i = 1000$, $\omega_1 = 3333$, $\omega_2 = 30000$, $\omega_f = 1e5\ rad/s$ and $K = 33.6$. Then, $S_c$ is as shown in Fig. \ref{fig:Example_1} with $|S_c| \geq 1$ for all frequencies greater than $1\ KHz$. Limiting peak of overall sensitivity function requires peak of $(1 - Q)$ to be below this frequency. From disturbance rejection perspective, ideal $Q$ design is with corner frequency at bandwidth which is also the worst case scenario for sensitivity peak. Consider choosing $Q$ for such a scenario with $\omega_q = 1e4\ rad/s$. $1 - Q$ and corresponding overall sensitivity function are also shown in Fig. \ref{fig:Example_1} and the result of two peaks coinciding can also be seen. 

From general requirements on $Q$ design, it can be seen that disturbance rejection property is compromised to attenuate measurement noise and vice-versa. Moreover, as seen from Fig. \ref{fig:Example_1}, with a linear low pass filter, design of $Q$ will need avoidance of coincidence of peaks which then requires the disturbance rejection to be typically limited to half of the uncompensated linear plant bandwidth. 

In general, the definitions and design guidelines for $Q$, stability conditions for overall compensated system and further analysis of linear DOB can be found in \cite{schrijver2002disturbance,umeno1991robust,shim2009almost}. While they have unbeatable simplicity, their effectiveness is limited due to noise attenuation - disturbance rejection trade-off and further by requirement of sensitivity peaks not coinciding. Generally, $Q$ can be kept close to unity up-to generally around half the uncompensated system bandwidth \cite{schrijver2002disturbance}. The higher harmonics generated by nonlinearities especially those occurring in the region of peak of $S_c$ stay uncompensated and this may even cause instabilities. Out of the two well-known ill-effects of nonlinearity namely imprecision and instability, only imprecision in the low frequency region is addressed. The limitations of DOB and its approach as listed out in this section are fundamental limitations of using linear techniques. Hence, the nonlinear reset control technique is considered to overcome these limitations with the required preliminaries for the novel RDOB configurations explained in the next section. 

\section{Reset control}
\label{Resetprel}

{
Reset is a nonlinear control where the states of the controller are reset under certain conditions. It was first introduced by Clegg in 1958 with an integrator whose state was reset when the error input was equal to zero \cite{clegg1958nonlinear}. Clegg also showed through describing function analysis that the reset integrator has a phase lag of only $38.1^\circ$ compared to linear integrator having $90^\circ$ which provides a $52.1^\circ$ phase advantage. Over the years, this theory has been greatly developed to generalize reset and use this phase advantage to increase robustness of closed-loop systems compared to its linear counterpart \cite{banos2011reset,guo2009frequency,hazeleger2016second,heertjes2015design,hosseinnia2013basic,hosseinnia2013fractional,hosseinnia2014general,saikumargeneralized,arun2018,linda2018}. One of the biggest advantages of reset is that describing function (although only an approximation of frequency domain behaviour) can be used for analysis and design. Over the past few years, reset control has also been popularly tested on precision motion control in \cite{li2011optimal, li2011reset,guo2009frequency,li2005nonlinear,li2009discrete}.
}

\subsection{Definition of reset elements}

Typically, reset elements are linear elements where the states are reset to zero when the input crosses zero. Although other forms of reset have been explored in literature, this is the most popular and we will restrict our discussion to this category. A reset element $C_R$ is expressed using differential inclusion as:

\begin{equation}
\label{GReset}
C_{R}= 
\begin{cases}
\dot{x_r}(t)=A_r x_r(t)+B_re(t) 			& \forall{e(t)\neq 0}\\
x_r(t^+)=A_\rho x_r(t) 					& \forall{e(t)= 0},\\
u_r(t)=C_rx_r(t)+D_re(t)
\end{cases}
\end{equation}

{
where $A_r$, $B_r$, $C_r$ and $D_r$ are the state matrices of the base linear system.  $x_r$ is the state vector, $e(t)$ is the input to the reset element and $u_r(t)$ is the output of the reset element. $A_\rho$ is the resetting matrix which determines the after-reset values of the states. Traditionally, this is a zero matrix or a combination of identity and zero matrices. From this definition, it can be seen that a reset element behaves mainly as a linear element with the first and third equation of \ref{GReset} operating till the input hits zero. At this point, the jump state is entered with second equation of \ref{GReset} where the states are reset based on composition of $A_\rho$.
}

\subsection{Describing function analysis}

Since reset is nonlinear, describing function is used in literature to obtain the approximate frequency behaviour. Sinusoidal input describing function analysis is provided in \cite{guo2009frequency} as 

\begin{equation}
C_R(j\omega)=C_r^T(j\omega I-A_r)^{-1}(I+j\Theta_\rho(\omega))B_r + D_r
\label{eq:tf}
\end{equation}
where   
$$
\Theta_\rho  =  \frac{2}{\pi}(I + e^{\frac{\pi A_r}{\omega}})\Big(\frac{I - A_\rho}{I + A_\rho e^{\frac{\pi A_r}{\omega}}}\Big)\Big(\Big(\frac{A_r}{\omega}\Big)^2 + I\Big)^{-1}
$$

The advantage of reset element compared to its linear counterpart is seen in reduced phase lag with describing function analysis as noted before. For example, consider a second order resetting low pass filter also known as Second Order Reset Element (SORE) which can be represented as

\begin{equation}
\label{Reset-lag}
R(s) = \dfrac{1}{\Big(\dfrac{\cancelto{}{s}}{\omega_r}\Big)^2 + \dfrac{2\zeta_r\cancelto{}{s}}{\omega_r} + 1}
\end{equation}

{
where $\omega_r$ is the corner frequency of LPF and $\zeta_r$ is the damping coefficient. The arrow indicates the resetting nature of SORE. The matrices for SORE for definition according to Eqn. \ref{GReset} are
}

\begin{equation*}
A_r=\begin{bmatrix}
0 				& 		1\\
-\omega_r^2 	&     -2\beta_r\omega_r
\end{bmatrix}
,
B_r=\begin{bmatrix}
0\\
\omega_r^2
\end{bmatrix}
\end{equation*}
\begin{equation*}
C_r=\begin{bmatrix}
1 & 0\\
\end{bmatrix},
D_r=\begin{bmatrix}
0
\end{bmatrix},
A_\rho = \emptyset_{2\times 2}
\end{equation*}

\begin{figure}[!htpb]
	\centering
	\includegraphics[trim = {2cm, 0, 2cm, 0}, width=\textwidth, keepaspectratio]{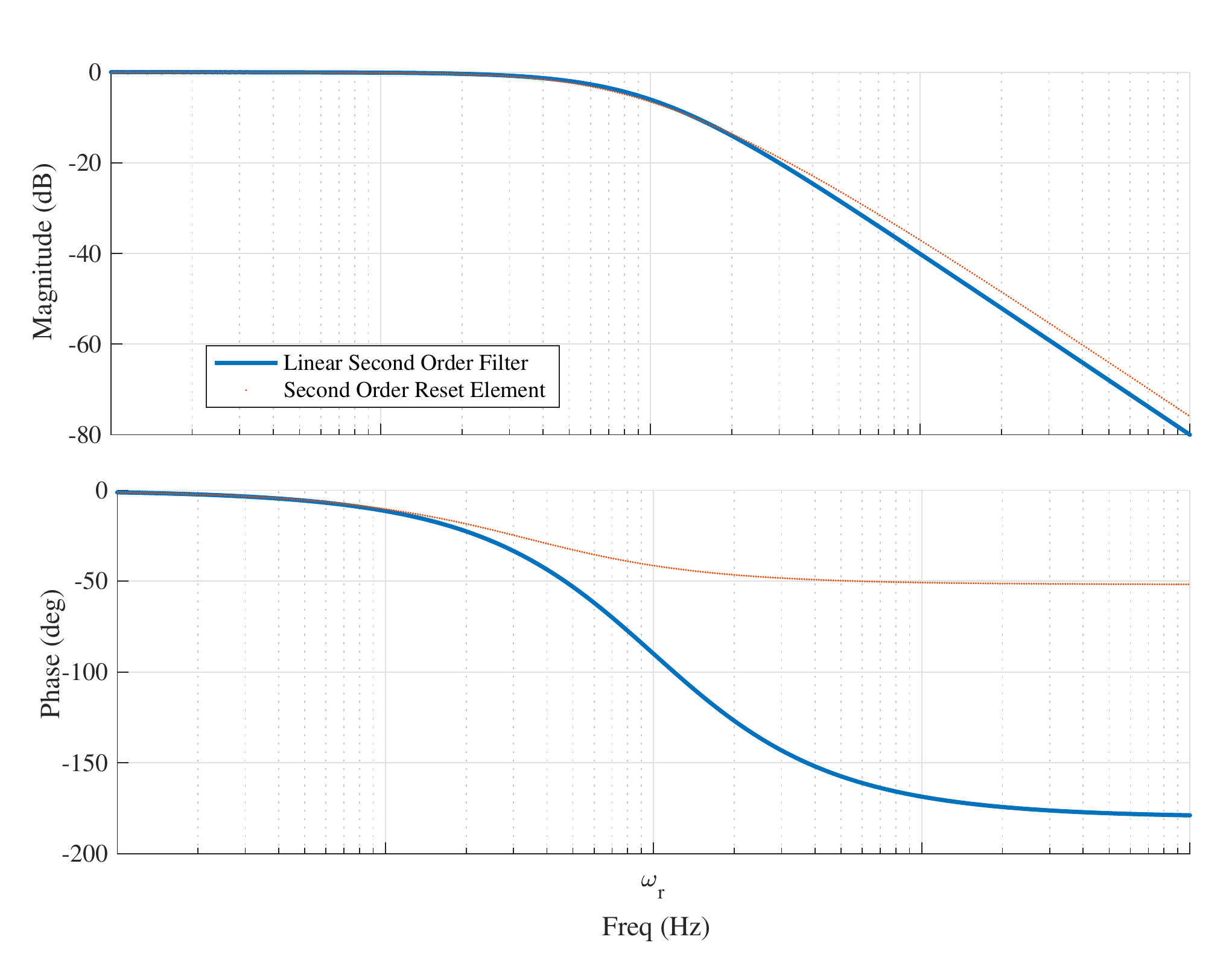}
	\caption{Frequency response of SORE obtained through describing function analysis (Eqn. \ref{eq:tf}) compared with frequency response of linear second order filter for $\zeta_r = 1$}
	\label{fig:GFORE}	
\end{figure}

The frequency response of SORE is obtained using Eqn. \ref{eq:tf} and shown in Fig. \ref{fig:GFORE} and compared against a linear second order LPF (for $\zeta_r = 1$). The shown response is true for any value of $\omega_r \in \mathbb{R}$. It is clear that SORE has $128.2^\circ$ lesser phase lag compared to linear one. While small differences are also seen in magnitude behaviour, these are quite small compared to the large phase lag advantage. The difference in magnitude is mainly seen as a small shift in corner frequency since the slope at higher frequencies is the same as that of linear LPF. The phase lag reduction and the small shift in magnitude are also seen with first order filters and integrators and the former has been used to great advantage.

\subsection{Stability of reset systems}

Two different stability criteria, one based on sinusoidal input and another based on Lyapunov function can be used to analyze stability of closed-loop systems with reset elements.

\subsubsection{Sinusoidal input response based stability}

Consider a system with reset element present in the loop. Open loop of such a system including plant can be represented as:

\begin{equation}
\label{stabreset}
\sigma_r= 
\begin{cases}
\dot{x}(t)=A x(t)+Br_u(t) 			& \forall{r_u(t)\neq 0}\\
x(t^+)=\overline{A_\rho} x(t) 					& \forall{r_u(t)= 0},\\
y(t)=Cx(t)+Dr_u(t)
\end{cases}
\end{equation}

where $r_u,y \in \mathbb{R}$ are the scalar input and output of the loop respectively, $x$ is the state vector. Since the loop consists of non-reset elements, $\overline{A_\rho}$ is not a zero matrix, but instead is a diagonal matrix with diagonal elements taking the form $[0,0...0,1,1,...1]$, where the zeros correspond to the states of reset element and ones correspond to states of non-reset elements (including plant). 

The stability of this loop is analysed using sinusoidal input as provided in \cite{guo2009frequency}. Let $r_u(t) = \alpha sin(\omega t)$. Thus the reset time instants {$t_k$} is given by
$$
t_k = k\pi/\omega,\ \ \ \ \ k = 0,1,..
$$

Defining $\eta_k = x({t_{2k}}^+), \zeta_k = x({t_{2k+1}}^+)$ and
$$
\psi(t) = \triangleq \int_{0}^{t}e^{-As}Bsin(\omega s)ds
$$

Then for the open loop system provided in Eqn. \ref{stabreset} with initial condition $x(0^+) = \nu_0$, we get

$$
x(t) = 
\begin{cases}
e^{A(t - t_{2k})}\eta_k + \alpha e^{At}[\psi(t) - \psi(t_{2k})],	&	t \in (t_{2k},t_{2k+1}]\\
e^{A(t - t_{2k+1})}\zeta_k + \alpha e^{At}[\psi(t) - \psi(t_{2k+1})],	&	t \in (t_{2k+1},t_{2k+2}]\\
\end{cases}
$$

where $\eta_k$ and $\zeta_k$ are obtained by the recursive algorithm,

$
\begin{cases}
\zeta_k = A_\rho e^{\frac{\pi}{\omega}A} [\eta_k + \alpha\psi(\dfrac{\pi}{\omega})],\\
\eta_k = A_\rho e^{\frac{\pi}{\omega}A} [\zeta_k - \alpha\psi(\dfrac{\pi}{\omega})],	&	\eta_0 = x(0+)
\end{cases}
$

This recursion converges globally if and only if

\begin{equation}
\label{resetstab}
|\lambda(A_\rho e^{A\frac{\pi}{\omega}})| < 1
\end{equation}

For full proof stability, see \cite{guo2009frequency}.

\subsubsection{Quadratic Lyapunov function based stability}

{
A simple condition based on transfer function matrix $H_\beta$ referred to as $H_\beta-condition$ is presented to check closed-loop stability of reset control systems in \cite{banos2011reset}. This condition is valid if states are reset when input is equal to zero (which is the case in this paper). The main results are provided below.
}

\begin{thm}{\cite{banos2011reset}.}\label{th:1}
	
	{
	Let $V: \mathbb{R}^n\rightarrow\mathbb{R}^n$ be a continuously differentiable, positive-definite, radially unbounded function such that
	}
	\begin{eqnarray}\label{eq:lyapunov}
	\dot{V}(x):=\Big(\pdv{V}{x}\Big)^TA_{cl}x<0,&\text{if } e(t) \neq 0,\\
	\Delta V(x):=V(A_\rho x)-V(x)\leq 0,& \text{if } \label{eq:lyapunov2} e(t)=0
	\end{eqnarray}
	
	{
	where $A_{cl}$ is the closed-loop $A$-matrix and $A_\rho$ is the reset matrix.
}
	
	{	
	Then the reset control system is asymptotically stable.
}
\end{thm}

{
Quadratic stability is guaranteed when (\ref{eq:lyapunov}) and (\ref{eq:lyapunov2}) hold true for a potential function $V(x)=x^TPx$ with $P>0$. From this condition the theorem to prove quadratic stability is obtained in \cite{banos2011reset} as follows.
}

\begin{thm}{\cite{banos2011reset}.} \label{th:2}
	
	{
	There exists a constant $\beta\in\mathbb{R}^{n_r\times 1}$ and $P_\rho\in\mathbb{R}^{n_r\times n_r}, P_\rho>0$ where $n_r$ is the number of reset states, such that the restricted Lyapunov equation
}
	\begin{eqnarray}\label{eq:lmi1}
	P>0,&A_{cl}^TP+PA_{cl}<0,\\
	&B_0^TP=C_0
	\end{eqnarray}
	
	{
	has a solution for $P$, where $C_0$ and $B_0$ are defined by:
}
	\begin{eqnarray}
	C_0=\begin{bmatrix}
	\beta C_{p}&O_{n_r\times n_{nr}}&P_\rho
	\end{bmatrix},
	B_0=\begin{bmatrix}
	O_{n_{p}\times n_r}\\O_{n_{nr}\times n_r}\\I_{n_r}
	\end{bmatrix}\label{eq:lmi2}
	\end{eqnarray}
	
	{
	and $n_{nr}$ is the number of non-reset states and $n_p$ is number of states in plant.
	}
\end{thm}

{
In both cases of stability provided, it is assumed that the base linear system is stable, i.e., if the jump condition is removed and reset element made linear, the overall closed-loop system is stable. The stability of such a linear system can easily be verified by linear control theory and is not dealt with here. In rest of the paper, sinusoidal input based stability condition is expanded upon to prove stability of the presented RDOB configurations.
}

\subsection{CgLp Element}

Most work in reset control relies on phase lag reduction used to improve phase margin and hence stability and robustness of closed-loop system. In \cite{gu2016modeling} and \cite{guo2009frequency} however, reset has been applied to a narrowband compensator to achieve phase lead and not just reduced phase lag.

Here, we introduce a novel reset element termed `Constant-gain Lead-phase' (CgLp) which provides phase lead over a broadband frequency range while maintaining a gain of one ($0\ dB$). This is achieved by utilizing a resetting low pass filter $R$ as described in Eqn. \ref{Reset-lag} in series with a non-resetting linear lead filter. The non-resetting linear lead filter $L$ can be described as

\begin{equation}
L(s) = \dfrac{(s/\omega_{r\alpha})^2 + (2s\zeta_r/\omega_{r\alpha}) + 1}{(s/\omega_f)^2 + (2s/\omega_f) + 1}
\label{Linear-lead}
\end{equation}
where $\omega_f >> \omega_{r\alpha}$ and $\omega_{r\alpha} = \alpha\omega_r$. $\omega_{r\alpha}$ determines the starting of lead action and $\omega_f$ indicates termination of the same. The reason for corner frequency defined to be at $\omega_{r\alpha}$ instead of $\omega_r$ is explained later.
If we consider asymptotic behaviour, $L$ provides phase lead in range $[\omega_{r\alpha},\omega_f]$. The series combination of $R$ and $L$ forms CgLp which can be represented in form of Eqn. \ref{GReset} with matrices

\begin{gather}
A_r=\begin{bmatrix}
0 				& 		1	&	0	&	0\\
-\omega_r^2 	&     -2\zeta_r\omega_r	&	0	&	0\\
0	&	0	&	0	&	1\\
1	&	0	&	-{\omega_f}^2	&	-2\omega_f
\end{bmatrix}
\\
B_r=\begin{bmatrix}
0\\
\omega_r^2\\
0\\
0
\end{bmatrix}\\
C_r=\begin{bmatrix}
\dfrac{{\omega_f}^2}{{\omega_{r\alpha}}^2}	&	0	&	{\omega_f}^2 - \dfrac{{\omega_f}^4}{{\omega_{r\alpha}}^2}	&	\dfrac{2{\omega_f}^2\zeta_r}{\omega_{r\alpha}} - \dfrac{2{\omega_f}^3}{{\omega_{r\alpha}}^2}\\
\end{bmatrix}\\
D_r=\begin{bmatrix}
0
\end{bmatrix}\\
A_\rho = \begin{bmatrix}
0	&	0	&	0	&	0\\
0	&	0	&	0	&	0\\
0	&	0	&	1	&	0\\
0	&	0	&	0	&	1\\
\end{bmatrix}
\label{Arhocglp}
\end{gather}

\begin{figure}[!htpb]
	\centering
	\includegraphics[trim = {3cm 1cm 3cm 1cm}, width=\linewidth, keepaspectratio]{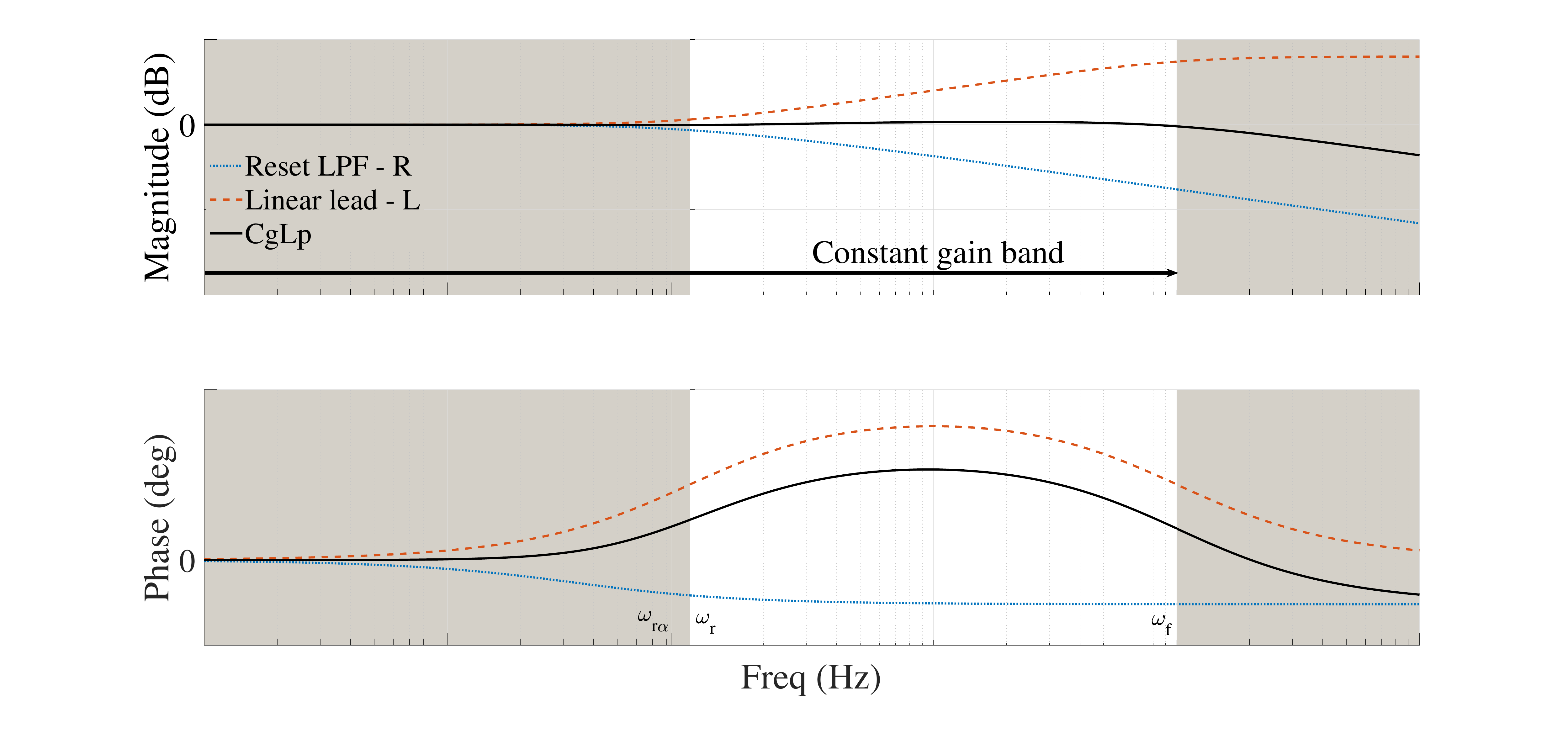}
	\caption{Frequency response of CgLp element obtained with SORE as filter $R$ through describing function analysis for $\omega_f = 100w_r$, $\zeta_r = 1$}
	\label{fig:CgLp}	
\end{figure}

In the linear case, having a linear LPF $R$ and linear lead $L$ (with $\alpha = 1$ resulting in $\omega_{r\alpha} = \omega_r$) will result in a second order LPF at $\omega_f$ due to cancellation. However, when $R$ is not linear and instead reset, phase lag is significantly reduced with a small effect on magnitude. This results in cancellation only in magnitude with constant gain ($0\ dB$), but phase lead over the broadband range $[\omega_r, \omega_f]$ as shown in Fig. \ref{fig:CgLp}. Depending on the range, up to $128.1^\circ$ phase lead can be obtained using this configuration of CgLp. This achieved phase lead with constant gain is true for all values of $\omega_r < \omega_{r\alpha} << \omega_f \in \mathbb{R}$. The reason for the use of $\alpha$ to offset the starting frequency of $L$ is to account for the small shift in corner frequency of $R$ as seen in Fig. Fig. \ref{fig:GFORE}. For SORE, $\alpha = 1.25$. 

While CgLp has been presented with second-order reset LPF (SORE) filter $R$ and second order lead filter $L$, this behaviour can also be achieved with first order elements. However, the use of second order has the advantage of achieving significantly higher phase lead compared to first-order \cite{hazeleger2016second}, along with the freedom to tune the damping factors of the filters. Hence this is considered and explained here. However, the fundamental concept of CgLp holds true even with the use of first order filters.

\section{Resetting Disturbance Observer}
\label{RDOB}

The broadband phase compensation achieved through CgLp is used as part of DOB resulting in Resetting DOB (RDOB). As discussed in Section. \ref{NPDOB}, linear DOB suffers from fundamental limitations of linear control. To overcome these, two novel configurations of RDOB which attempt to solve the problem from two different perspectives are presented.

\subsubsection{Direct phase compensation RDOB : config - 1}

As explained in Section. \ref{NPDOB}, the performance of linear DOB based compensation architecture heavily depends on the design of disturbance estimating $Q$. To avoid coinciding sensitivity peaks, $Q$ is generally designed such that its action is limited to around half of uncompensated closed-loop bandwidth. However, even in the case where this is ignored and $Q$ is designed to operate till the bandwidth, disturbance rejection property of this compensation scheme is reduced and compromised close to the corner frequency of $Q$-filter due to the introduced phase lag which is the natural property of a linear low pass filter. To overcome this fundamental limitation, we propose `Direct phase compensation RDOB' where CgLp is used in series with $Q$. Since CgLp is designed to have constant gain behaviour, the low pass filtering effect of $Q$ in inner loop is maintained. However, since CgLp adds phase, the overall phase lag of the inner loop is reduced significantly.

\begin{figure}[!htpb]
	\centering
	\includegraphics[width=1\linewidth]{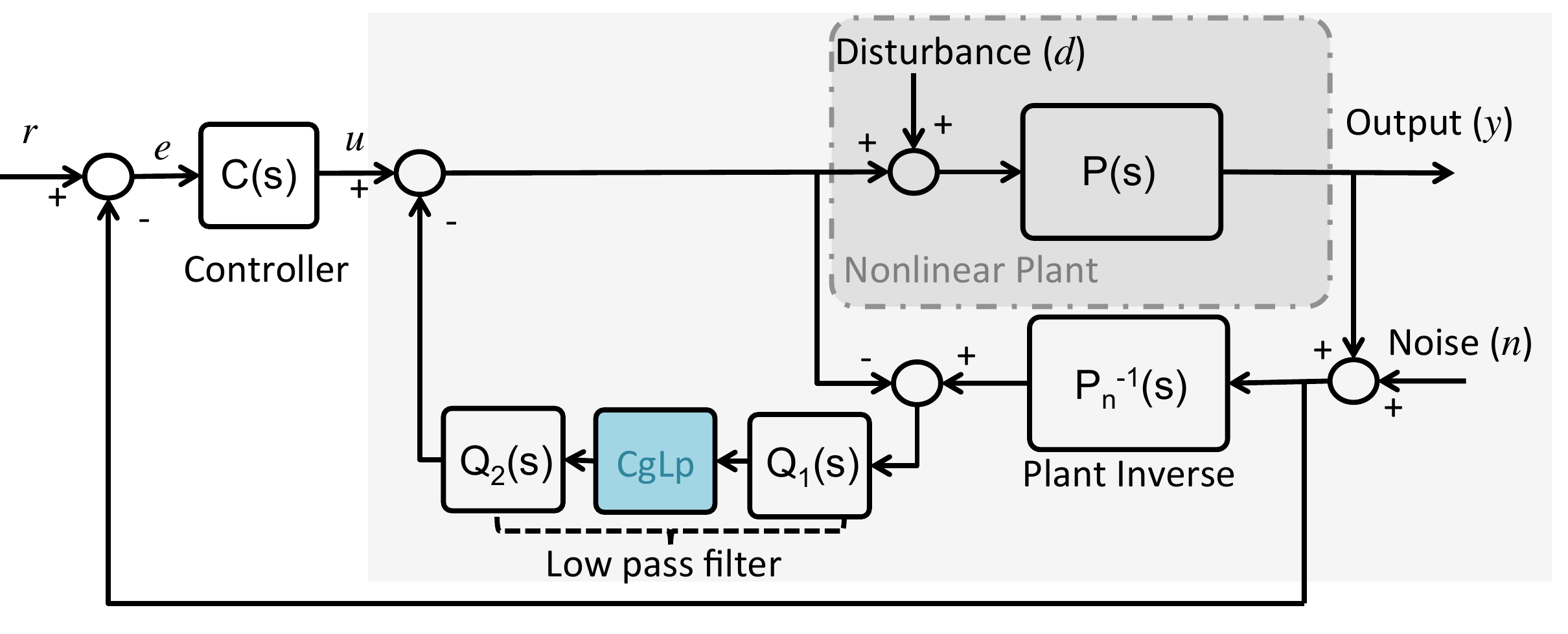}
	\caption{Direct phase compensation RDOB config-1}
	\label{fig:DOB+reset}	
\end{figure}

{
The block diagram of 'Direct phase compensation RDOB' (referred to as RDOB config-1) is shown in Fig. \ref{fig:DOB+reset}. The difference between RDOB config-1 scheme and existing linear DOB architecture of Fig. \ref{fig:DOB+cont} is that the new nonlinear disturbance estimating filter $Q$ is given as
}
\begin{equation}
Q(j\omega) = Q_1(j\omega).CgLp(j\omega).Q_2(j\omega)
\label{DOB1Qfilter}
\end{equation}

{
where $Q_1$ and $Q_2$ are linear filters and frequency domain behaviour of CgLp can be obtained through describing function analysis. The reason for using two filters $Q_1$ and $Q_2$ is explained later.
}

{
For the design of RDOB config-1, optimal design of the linear $Q$ is obtained as before from the outline in \cite{shim2009almost}. However, in this case, we design $Q$ to operate till the bandwidth and not half as is generally the case. It can be seen from Fig. \ref{fig:DOB+reset} that $Q$ has been split into 2 filters $Q_1$ \& $Q_2$ and CgLp strategically placed between them. This is done to avoid multiple resets of reset element CgLp due to measurement noise. $Q_1$ filters out this noise and $Q_2$ is designed such that it compliments $Q_1$. The additional advantage of this scheme is that $Q_2$ also filters out high-frequency harmonics arising from the resetting action. It should be noted again that the split in $Q$ does not compromise the disturbance estimating function of $Q$ since $Q_2$ is obtained such that the relation (\ref{div}) is satisfied, using originally designed $Q$ and $Q_1$ which is determined based on measurement noise present in the system. 
}

\begin{equation}
Q=Q_1\times Q_2
\label{div}
\end{equation}

{
Although RDOB config-1 is proposed to increase the effectiveness of disturbance rejection close to corner frequency of $Q$ (which is now designed to be at bandwidth), the advantage can also be seen in the sensitivity function. Phase lead of CgLp in RDOB config-1 improves the sensitivity function of inner loop $S$. This, in turn, results in an improvement in overall sensitivity. Hence, the problem of coincidence of sensitivity peaks is also indirectly tackled. This advantage is illustrated below using the plant which was considered in Section. \ref{NPDOB}.
}

For the simple second order plant considered before in (\ref{eq_pn}), consider the $Q$-filter as given by (\ref{eq_q}). This is split into two first order filters $Q_1$ and $Q_2$ for RDOB config-1. Also, CgLp is designed with $\zeta_r = 1$ and $\omega_r = \omega_q$.

\begin{figure}[!htpb]
	\centering
	\includegraphics[width=1\linewidth]{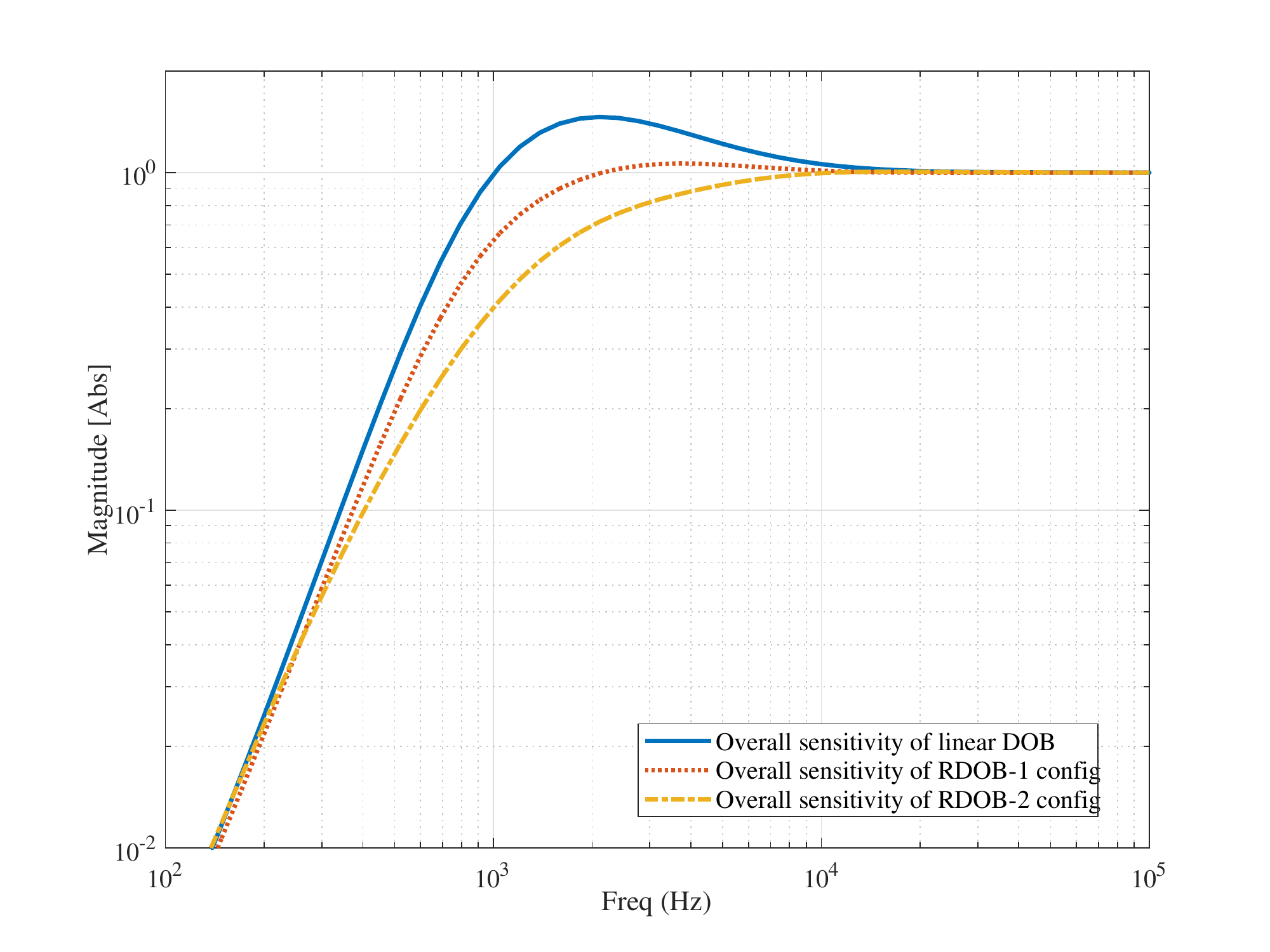}
	\caption{Overall sensitivity function for linear DOB, RDOB config-1 and RDOB config-2 obtained for introductory example}
	\label{fig:Example_4}	
\end{figure}

The overall sensitivity for RDOB config-1 for values as used in Section. \ref{NPDOB} is shown in Fig. \ref{fig:Example_4} along with that of the system with linear DOB and this shows the additional advantage of decrease in sensitivity peak with RDOB config-1. Although $\omega_r$ is chosen to be equal to $\omega_q$ for illustrative purposes, $\omega_r < \omega_q$ is a better choice to ensure greater phase compensation close to $\omega_q$. Ideally, $\omega_r$ can be chosen such that it completely compensates for phase lag of $Q$ at $\omega_q$, so that the combination of CgLp and $Q$-filter has zero phase at $\omega_q$. Also, a second order $Q$ has been considered here which has a phase lag of $90^\circ$ at corner frequency. However, when higher order $Q$-filters are designed, CgLp with second-order reset element will not be able to completely compensate the phase. In such cases, CgLp with higher order reset elements will have to be considered if complete phase compensation is desired. However, this is beyond the scope of this paper.

One of the main limitations on design of $Q$ is the contradictory requirements for disturbance rejection and noise attenuation. Although RDOB-1 continues to use the original LPF design for $Q$, this action is improved through the use of CgLp since low pass filtering action is maintained in gain, but phase lag problem is tackled through CgLp. This additionally reduces sensitivity peak of inner loop allowing for $Q$ to be designed at bandwidth instead of half of same.

\subsubsection{Controller side phase compensation RDOB : config - 2} 

\begin{figure}[!htpb]
	\centering
	\includegraphics[width=1\linewidth]{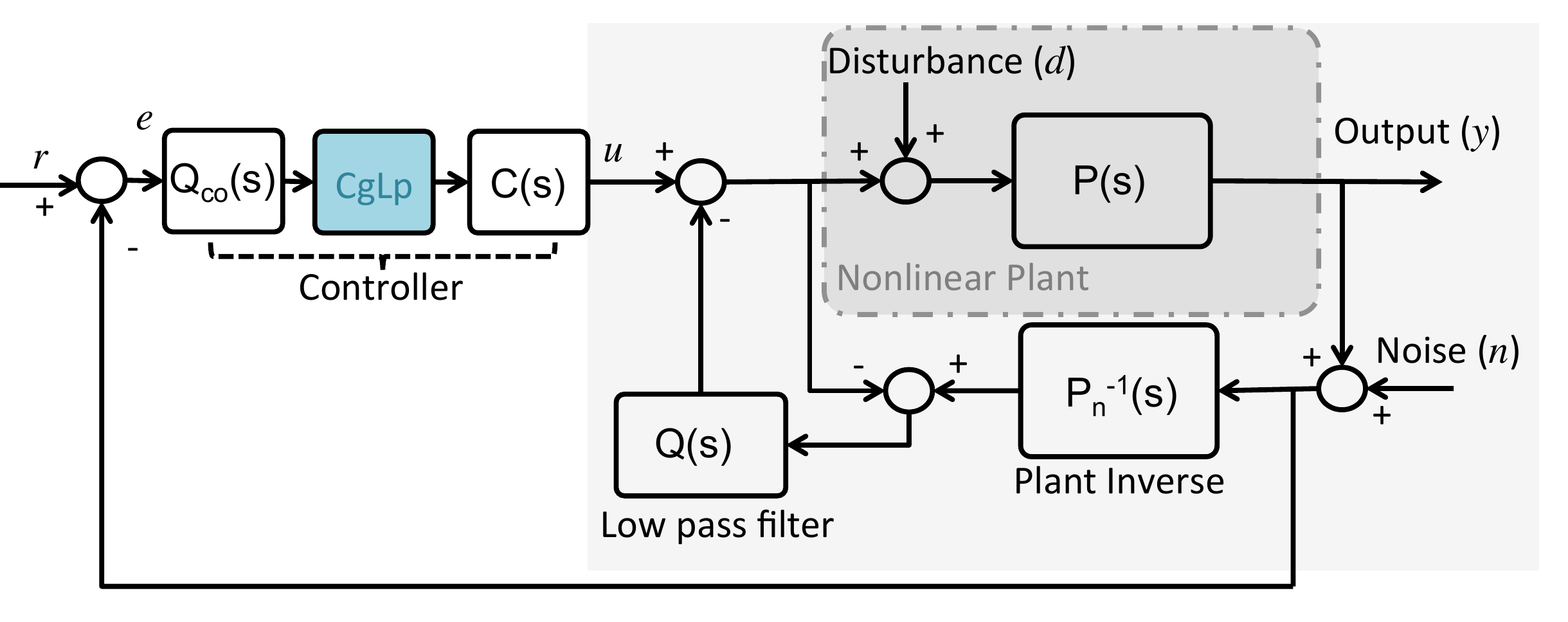}
	\caption{Controller based RDOB config-2}
	\label{fig:DOB+reset+controller}	
\end{figure}

RDOB config-1 tackles the problem of linear DOB by improving the inner loop. However, in this approach, sensitivity of outer loop $S_c$ remains the same. The coincidence of sensitivity peaks resulting in `inseparability of DOB and controller design' as explained in Section. \ref{NPDOB}, can also be tackled by directly attempting to reduce the sensitivity peaks instead of avoiding their coincidence. Here, a second resetting DOB configuration is proposed where CgLp is used in the outer loop similar to the idea proposed in \cite{li2011reset} and linear DOB is used in the inner loop without CgLp. The block diagram of this `Controller based RDOB' is shown in Fig. \ref{fig:DOB+reset+controller}.

{
The difference between RDOB config-2 scheme and existing linear DOB architecture of Fig. \ref{fig:DOB+cont} is that the new nonlinear feedback controller is given as
}
$$
C(j\omega) = Q_{co}(j\omega).CgLp(j\omega).C(j\omega)
$$

{
where $C$ is the same feedback controller designed for linear DOB and frequency response of CgLp can be obtained through describing function analysis. An additional filter $Q_{co}$ is added to filter measurement noise and avoid repetitive resets of CgLp. 
}

{
The filtering of measurement noise to avoid multiple resets was achieved by splitting $Q$ into $Q_1$ and $Q_2$ in RDOB config-1. Since an additional filter $Q_{co}$ is used for this purpose with RDOB config-2, additional phase lag is introduced. However, phase lead obtained from CgLp is larger and this enhances the stability margins of outer feedback control loop which in turn reduces the sensitivity peak of $S_c$ and hence reduces the overall sensitivity. In practice, an LPF at higher frequencies is already present in conventional controller design to attenuate high-frequency noise. Hence, in reality, this filter can be used as $Q_{co}$ and ensure that this phase loss is also not seen. In either case, the reduction seen in peak of sensitivity through use of CgLp is again illustrated through the example plant considered before.
}

With RDOB config-2, CgLp is used in series with the plant and controller in the outer loop. Similar to the example considered for config-1 , ${\zeta_r} = 1$ is chosen. However, here $\omega_r$ is considered equal to that of the plant ($w_p$. See Eqn. \ref{eq_pn}). The lead filter provides a phase lead from the same $w_p$ till $w_f >> w_p$. PID controller design in Eqn. \ref{eq_c} consists of a LPF at $\omega_f$ and this is used as $Q_{co}$.

The overall sensitivity is shown in Fig. \ref{fig:Example_4} for values considered in Section. \ref{NPDOB}. It is seen that RDOB config-2 provides even better performance than config-1 with $|S\times S_c| <= 1$ for all frequencies. 

While RDOB config-1 attempts to solve the first limitation on $Q$ design, RDOB config-2 attempts to tackle the second limitation by directly reducing the sensitivity peak of uncompensated system $S_c$. This not only solves the problem of coinciding sensitivity peaks but as seen from example plant considered, $|S\times S_c| <= 1$ (at least as seen for this plant) can result in significantly improved performance.

\subsection{Stability Criteria}

Sinusoidal input-based and Quadratic Lyapunov based stability conditions have been presented for reset control systems. Here, we use the first condition to provide stability matrices for presented RDOB configurations. The augmented systems shown in Fig. \ref{fig:DOB} and Fig. \ref{fig:DOB+cont} are used for stability analysis. 

\subsubsection{Stability of RDOB config-1}

For RDOB config - 1, CgLp is added into the disturbance estimating part with $Q$ split into $Q_1$ and $Q_2$ as shown in fig. \ref{fig:DOB+reset}. The open loop state matrices of this new augmented system which can be represented by Eqns. \ref{aug1} and \ref{aug2} are as given in Eqns. \ref{A_2} and \ref{BC_2}.

\footnotesize
\arraycolsep=3pt 
\begin{figure*}[h]
	\begin{gather}
	A=\begin{bmatrix}
	A_{P} 					& 0 				 & 	0	          & 0		  &-B_PC_{Q_2}\\
	B_{P_n^{-1}}C_P 		& A_{P_n^{-1}}		 &	0	          & 0			& 0		  \\
	B_{Q_1}D_{P_n^{-1}}C_P	& B_{Q_1}C_{P_n^{-1}}& 	A_{Q_1}   	  & 0	 	 & B_{Q_1}C_{Q_2} \\
	0 						& 0					 &	B_{CgLp}C_{Q_1}	  & A_{CgLp}			& 0		  \\
	0 						& 0					 &	B_{Q_2}D_{CgLp}C_{Q_1} & B_{Q_2}C_{CgLp}		& A_{Q_2}	  \\
	\end{bmatrix}
	\label{A_2}\\
	B_1=\begin{bmatrix}
	B_P\\
	0\\
	-B_{Q_1}\\
	0\\
	0
	\end{bmatrix}
	B_2=\begin{bmatrix}
	0\\
	B_{P_n^{-1}}\\
	B_{Q_1}D_{P_n^{-1}}\\
	0\\
	0
	\end{bmatrix},	
	C=\begin{bmatrix}
	C_P & 0 & 0 & 0 & 0
	\end{bmatrix}
	\label{BC_2}	
	\end{gather}
	where, $x(t)$=[$x_P(t)$ $x_{P_n^{-1}}(t)$ $x_{Q_1}(t)$ $x_{CgLp}(t)$ $x_{Q_2}(t)$]$^T$ and, \\
	$A_P, B_P, C_P$ are the state matrices of plant $P$,\\
	$A_{{P_n}^{-1}}, B_{{P_n}^{-1}}, C_{{P_n}^{-1}}, D_{{P_n}^{-1}}$ are the state matrices of estimated plant inverse ${P_n}^{-1}$,\\
	$A_{Q_1}, B_{Q_1}, C_{Q_1}$ are the state matrices of filter $Q_1$,\\
	$A_{Q_2}, B_{Q_2}, C_{Q_2}$ are the state matrices of $Q_2$,\\
	$A_{CgLp}, B_{CgLp}, C_{CgLp},D_{CgLp}$ are the state matrices of CgLp element such that they can be represented as		
	\begin{gather}
	\label{aug1}
	\dot{x}(t)=A x(t)+B_1u(t)+B_2n		\\
	\label{aug2}
	y(t)=Cx(t)			
	\end{gather}
\end{figure*}
\normalsize
For this configuration, $A_\rho$ is a diagonal matrix with ones except for the first two states of CgLp as shown in Eqn. \ref{Arhocglp}. The matrices $A$ and $A_\rho$ as defined above for RDOB config-1 can be used in Eqn. \ref{resetstab} to analyse stability.

\subsubsection{Stability of RDOB config-2}

In the case of RDOB config - 2, CgLp is placed in series with the controller. Hence, the augmented system in this case as shown in Fig. \ref{fig:DOB+reset+controller} has to include the controller $C$ and is given by set of equations
\begin{gather}
\dot{x}(t)=A x(t)+B_1r(t)+B_2n		\\
y(t)=Cx(t)			
\end{gather}
where $r(t)$ is the reference tracking signal.
The state matrices of this system are given in Eqns. (\ref{A_3}) and (\ref{BC_3}). These state matrices are provided assuming that a separate $Q_{co}$ is designed and LPF which is part of controller is not used.

\footnotesize
\arraycolsep=3pt 
\begin{figure*}[h]
	\begin{gather}
	\scriptsize{
	A=\begin{bmatrix}
	A_{P} 					& 0 				 & 	-B_PC_{Q_2}	          & 0	  	      &B_PD_CC_{CgLp}   	&B_PC_C\\
	B_{P_n^{-1}}C_P 		& A_{P_n^{-1}}		 &	0	          		  & 0	  	      & 0		  	&0\\
	B_{Q}D_{P_n^{-1}}C_P	& B_{Q}C_{P_n^{-1}}	 & 	A_{Q}+B_QC_Q   	  	  & 0	  	      & -B_QD_C	&-B_QC_CC_{CgLp}\\
	0 						& 0					 &	0	  				  & A_{Qco}	      & 0	   	&0\\
	0 						& 0					 &	0					  & B_{CgLp}C_{Qco} 	  & A_{CgLp}		&0\\
	0 						& 0					 &	0	      	          & B_CD_{CgLp}C_{Qco}	  & B_CC_{CgLp} 	&A_C\\
	\end{bmatrix}}
	\label{A_3}\\
	B_1=\begin{bmatrix}
	0\\
	0\\
	0\\
	B_{Q_{co}}\\
	0\\
	0
	\end{bmatrix},
	B_2=\begin{bmatrix}
	0\\
	B_{P_n^{-1}}\\
	D_{P_n^{-1}}B_Q\\
	0\\
	0\\
	0
	\end{bmatrix},	
	C=\begin{bmatrix}
	C_P & 0 & 0 & 0 & 0 & 0
	\end{bmatrix}
	\label{BC_3}
	\end{gather}
	where, $x(t)$=[$x_P(t)$ $x_{P_n^{-1}}(t)$ $x_Q$ $x_{Q_{co}}$ $x_{CgLp}$ $x_C(t)$]$^T$ and, \\
	$A_C, B_C, C_C, D_C$ are the state matrices of linear controller such that they can be represented as in Eqns. \ref{aug1} and \ref{aug2}.
\end{figure*}
\normalsize

As described in \cite{guo2009frequency}, reset systems of both configurations have a global asymptotically stable $2\pi/\omega$-periodic solution under sinusoidal input for any aribitrary frequency $\omega > 0$ if and only if Eqn. \ref{stabreset} is satisfied $\forall\ \omega \in [0,\infty)$.

\section{Application in Piezo-Actuator}
\label{PZ}

\begin{figure}[!htpb]
	\centering
	\includegraphics[height = 0.6\textheight]{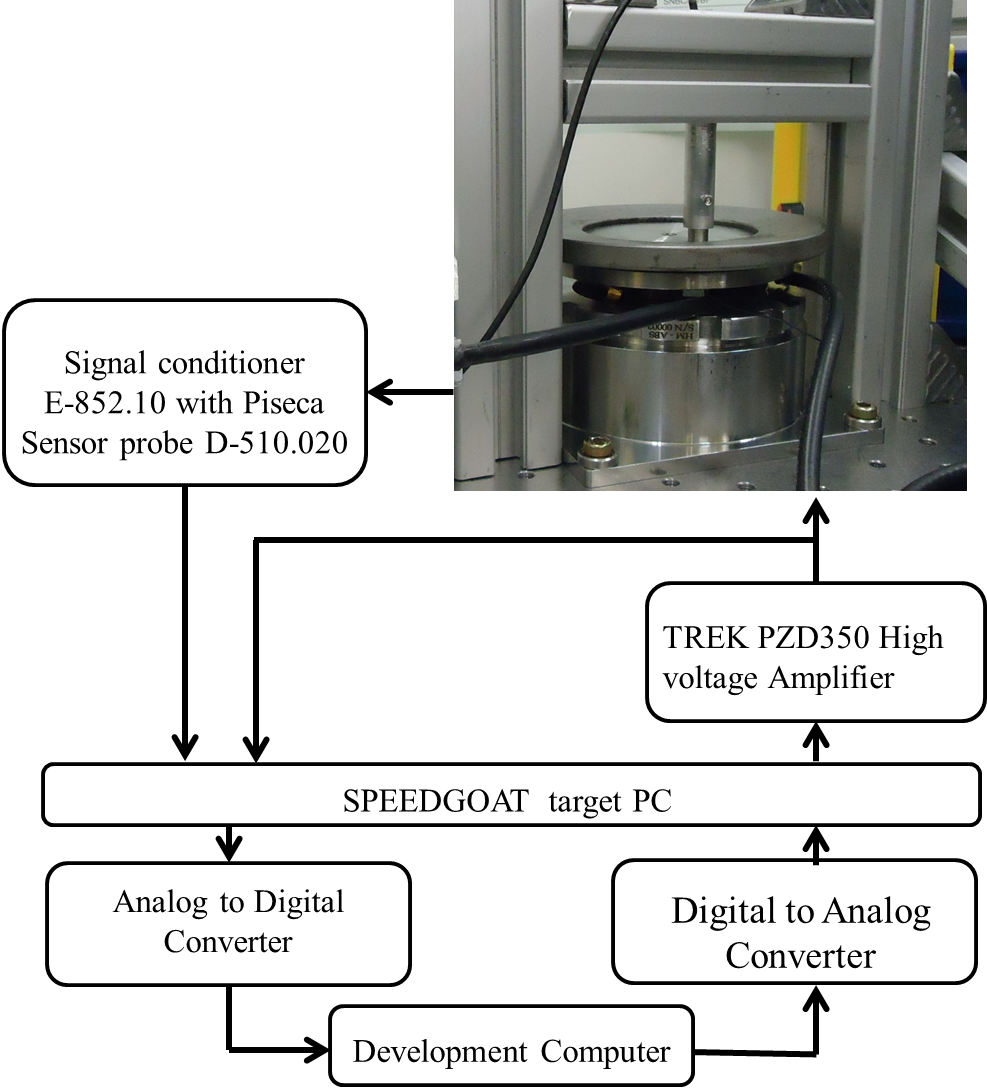}
	\caption{Experimental Setup}
	\label{fig:expt_setup}	
\end{figure}

The presented novel RDOB configurations are considered for hysteresis compensation on a piezo-actuator where its performance is compared against that of a linear DOB for validation of presented concepts. The experimental set-up is shown in Fig. \ref{fig:expt_setup}.

\subsection{Piezo-Actuator}

A piezo-stack is used for this purpose with a mass attached at one end and the other end fixed. Piezo-layers ($d_{33} = 10^{-8}m/V$) are arranged with the orientation to ensure mechanical deformation of all layers in the same direction. A voltage amplifier ('TREK PZD350') is used to actuate and a capacitive sensor ('PIseca capacitive sensing probe D-510.020 with signal conditioner E-852.10') having sensitivity of $0.2 V/ \mu m$ and static resolution of $0.5 nm$ is used to measure the displacement of free-end of the piezo-actuator. 

The nominal plant model is estimated at a sampling rate of $10 KHz$ (with the same rate used later for control) and found to be:
\begin{equation}
\label{plantnominal}
P_n=\frac{5.8\times10^4(s^2+439.8s+1.934\times 10^7)}{(s^2+754s+1.421\times10^7)(s^2+638.3s+3.948\times10^7)}
\end{equation}

\begin{figure}[!htpb]
	\centering
	\includegraphics[width=1\linewidth]{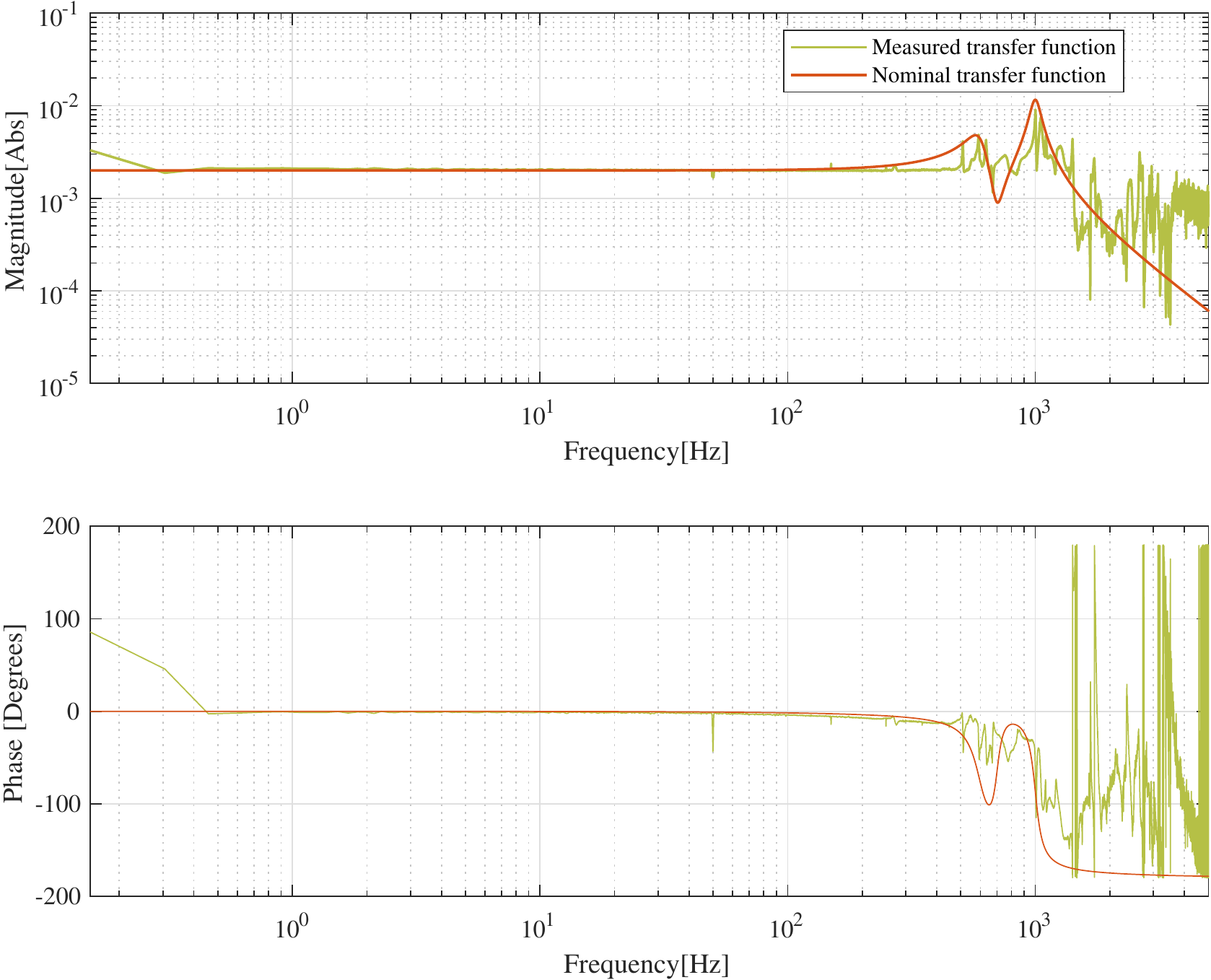}
	\caption{Bode Plot of the measured and nominal plant}
	\label{fig:bode_plant}	
\end{figure}

The measured frequency response and bode plot nominal transfer function from input voltage to the displacement are shown in Fig. \ref{fig:bode_plant}. The nominal plant model of Eqn. \ref{plantnominal} is estimated by considering only the first two eigenmodes of the plant and this will be used as part of the different DOB approaches especially for stability analysis.

\subsection{Controller design}

An $H_\infty$ controller is designed to be used in the outer loop using LMI (linear matrix inequalities) based technique such that the transfer of noise to the output is minimized. It is designed using the SEDUMI solver within the YALMIP environment for the nominal plant estimated in Eqn. \ref{plantnominal}. A sub-optimal controller is obtained using a typical weighting function ($W$) given as:
\begin{equation}
W=\frac{s^2+1885s+3.553\times10^6}{s^2+119.22s+3.553\times10^3}
\end{equation}

The obtained $H_\infty$ controller is given in Eqn. \ref{eq_controller1}. 

\begin{figure*}[!htpb]
	\vspace*{4pt}
	\hrulefill
	\footnotesize{
	\begin{equation}
	C =\frac{777.9\times(z-0.9725)(z+0.06267)(z^2-1.844z+0.9391)(z^2-1.523z+0.8819)}{(z+0.904)(z-0.9979)(z-0.9982)(z+0.02331)(z^2-1.555z+0.789)}
	\label{eq_controller1}
	\end{equation}}
	\hrulefill
	\vspace*{4pt}
\end{figure*}

\subsection{Design of the disturbance estimating $Q$ filter}

The nominal plant has a relative order of two, requiring the relative order of $Q$ to be two or higher to make the inner loop realizable. This is chosen as two and $Q$ designed using the method given in \cite{shim2009almost}. The obtained $Q$ is given in Eqn. \ref{Q-filter}.
\begin{equation}
Q=\frac{2.31\times10^{-3}(z+3.104)(z+0.221)}{(z-0.5762)(z^2-1.789z+0.8165)}
\label{Q-filter}
\end{equation}

\begin{figure*}
	\centering
	\includegraphics[width=1\linewidth]{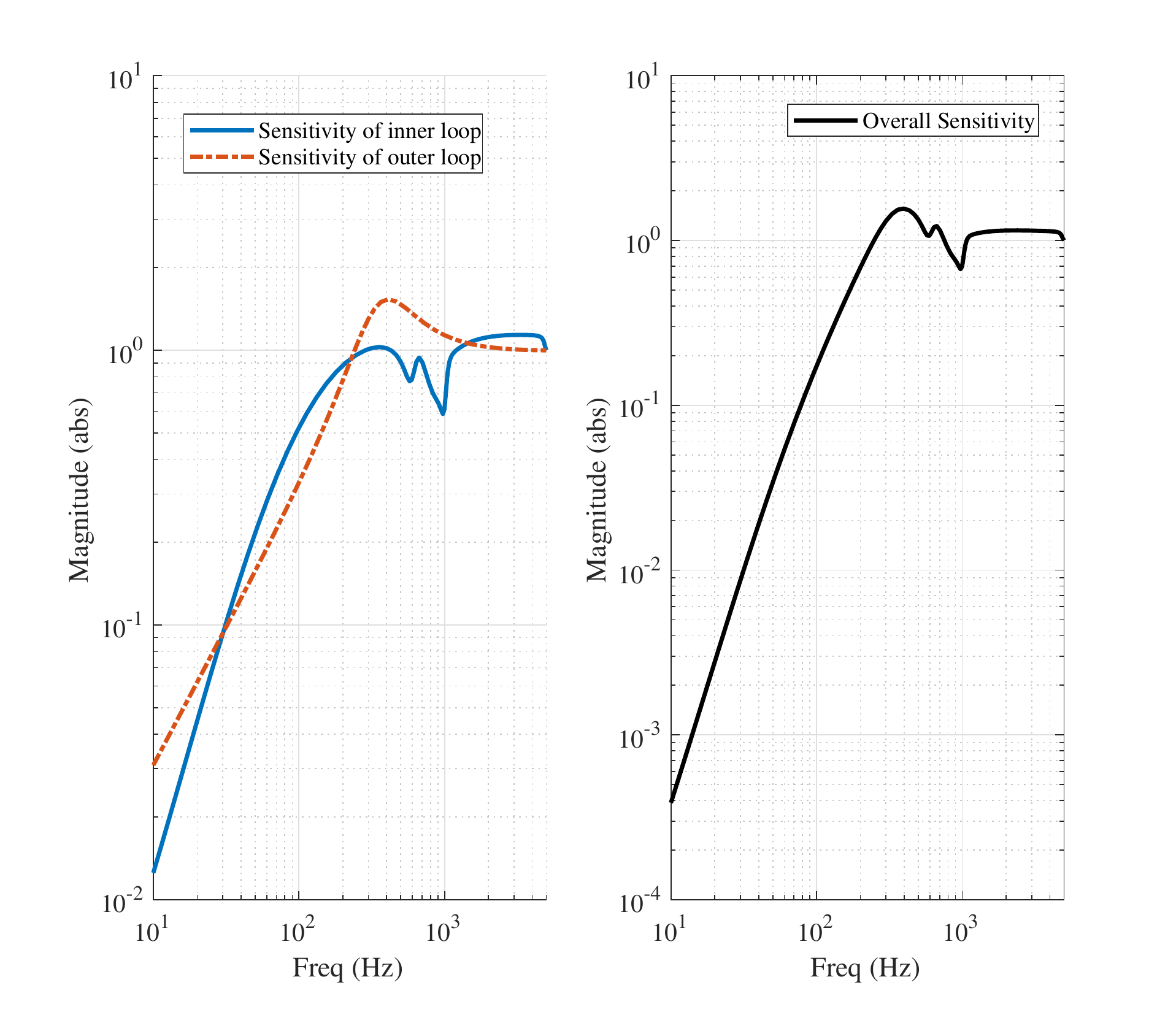}
	\caption{Sensitivity Function of the inner compensation scheme and the outer controlled plant}
	\label{fig:sensi}	
\end{figure*}

The theoretical sensitivity functions of DOB and controlled plant are shown in Fig. \ref{fig:sensi}. It can be seen that $Q$ has been designed to work at the maximum possible bandwidth with the peaks of sensitivity slightly coinciding. 

\subsection{Design of CgLp element}
\label{CgLpdesign}

Under config-1, CgLp is used to add phase to $Q$ close to its corner frequency to compensate for delay introduced by phase lag. Since, $Q$ corner frequency is approximately $228\ Hz$, CgLp is designed as per Eqns. \ref{Reset-lag} and \ref{Linear-lead} with $\omega_r = 2\pi 100$, $\omega_f = 2\pi 5000$ (where $5000\ Hz$ is half the sampling frequency) and $\zeta_r = 0.7$. With this choice of $\omega_r$ it is seen that CgLp compensates by almost $79^\circ$ in phase at $\omega_q$, resulting in just $11^\circ$ phase lag at the same frequency instead of $90^\circ$. 

Further, config-1 also involves splitting of $Q$ to $Q_1$ and $Q_2$. $Q_1$ is designed to attenuate measurement noise with corner frequency of $250 Hz$ as

\begin{equation}
Q_1=\frac{0.5302z^2 - 0.9712z + 0.4531}{z^2 - 1.881z + 0.8931}
\label{Qfilter}
\end{equation}
$Q_2$ is designed such that Eqn. \ref{div} is satisfied.

Under config-2, the idea is to add phase in the outer loop to reduce peak of sensitivity $S_c$. From Fig. \ref{fig:sensi}, the peak in outer loop is seen around $500\ Hz$. Hence, in this case, CgLp is designed to add phase around this region with $\omega_r = 2\pi 700$, $\omega_f = 2\pi 5000$ and $\zeta_r = 0.7$. Config-2 also involves the design of an additional filter $Q_{co}$ to filter out noise from the feedback signal used for resetting. LPF which is part of $H_\infty$ controller is used for this purpose to avoid additional phase lag.

\subsection{title}

\subsection{Stability of RDOB configurations}

\begin{figure}
	\centering
	\includegraphics[width=1\linewidth]{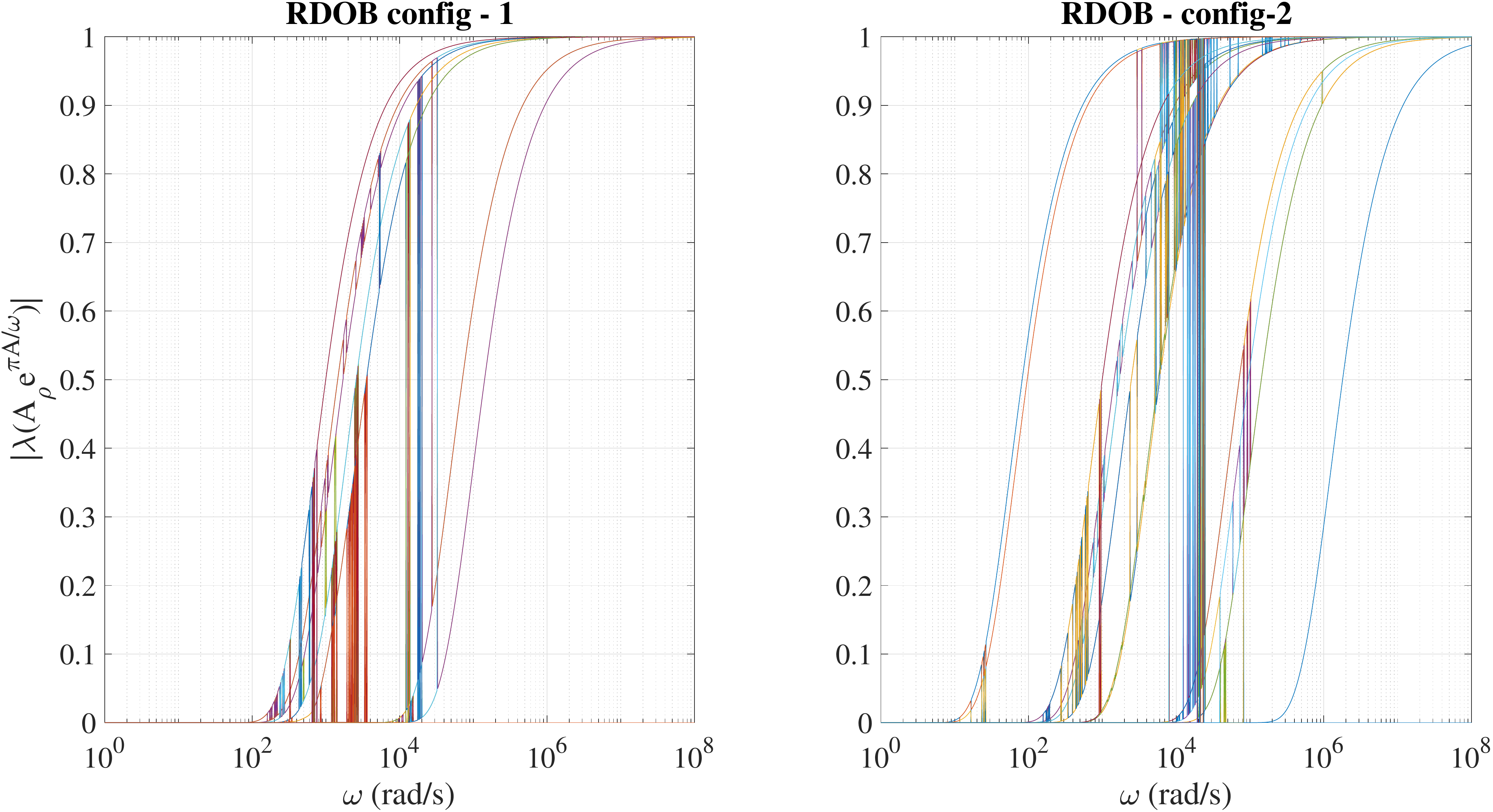}
	\caption{Magnitude of eigenvalues in open loop for both configurations of RDOB}
	\label{fig:stabplots}	
\end{figure}

The values of the designed elements are plugged into the state matrices of Eqns. \ref{A_2} and \ref{A_3} and eigenvalues found for a range of frequencies to ensure stability according to Eqn. \ref{stabreset}. The magnitude of eigenvalues is plotted in Fig. \ref{fig:stabplots}.

The eigenvalues satisfy the stability condition. The jump in eigenvalues seen in plots is due to MATLAB arranging eigenvectors in ascending order and outputting the corresponding eigenvalues. Since these eigenvalues are only used for confirmation of stability, the jumps seen in plots is not important.

\subsection{Results}

\begin{figure}
	\centering
	\includegraphics[width=1\linewidth,height=0.7\linewidth]{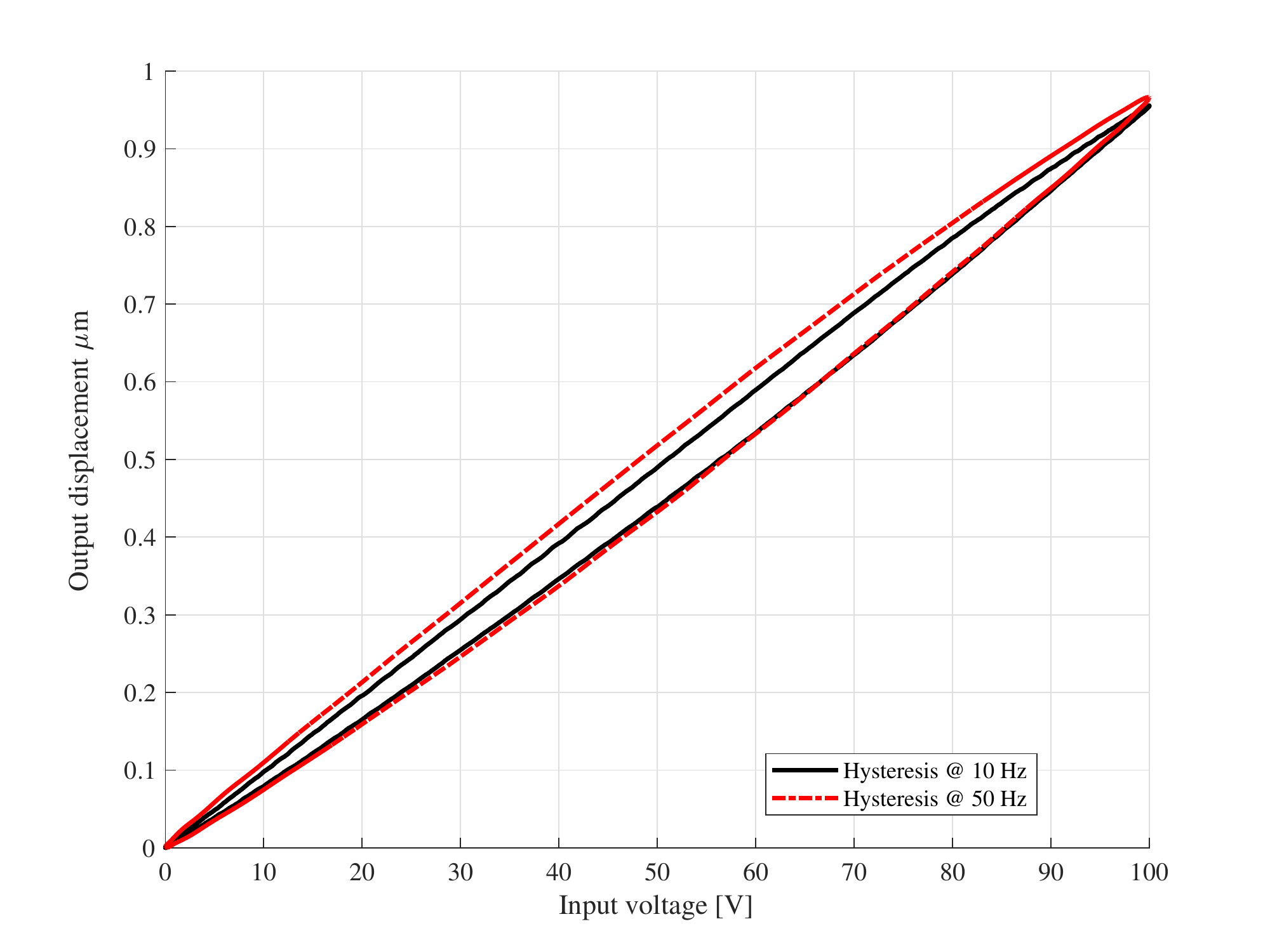}
	\caption{Identified Hysteresis in the piezo-actuator at $10\ Hz$ and $50\ Hz$}
	\label{fig:hyst1}	
\end{figure}

The piezo-actuator is actuated with a sinusoidal signal of $10\ Hz$ and also $50\ Hz$ sequentially. The plot between applied input voltage and output displacement as measured by a capacitive sensor is shown in Fig. \ref{fig:hyst1}. It can be seen that the piezo not only has asymmetric but also frequency dependent hysteresis.

\begin{figure}
	\centering
	\includegraphics[width=1\linewidth,height=0.7\linewidth]{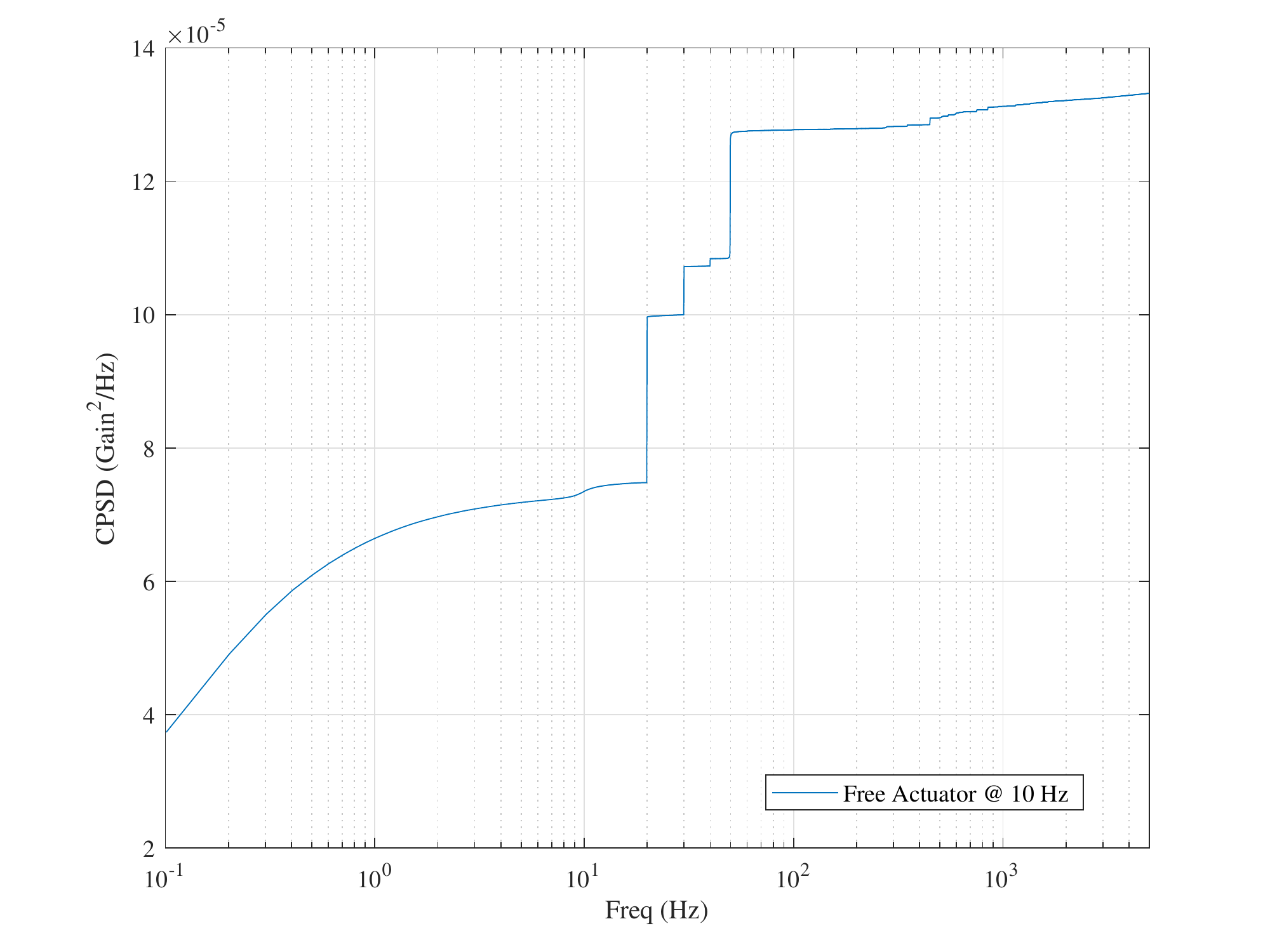}
	\caption{CPSD for the free actuator (in open loop) excited at $10\ Hz$}
	\label{fig:CPSD1}	
\end{figure}

{
To get an idea of this behaviour in frequency domain, cumulative power spectral density (CPSD) of open-loop error for $10\ Hz$ actuation is shown in Fig. \ref{fig:CPSD1}. This shows the dominant nonlinear hysteretic relation between voltage and displacement with higher order harmonics seen at integer multiples of $10\ Hz$. It is also noticed that some cumulative power is present till $10\ Hz$. We suspect that this is due to inherent effects in piezo-actuators like creep and also due to effects from impurities, humidity etc. However, we do not intend to use the actuator in open-loop and study of piezo-actuators is not the focus but instead on overcoming fundamental limitations of linear DOB. So these are not elaborated upon further.
}

A sinusoidal reference input of $30\ Hz$ is provided for tracking to all the different DOB compensation schemes including the two novel proposed configurations, and error between reference and actual position of piezo-actuator in terms of noise and hysteresis compensation is used for performance analysis.

\subsubsection{$H_\infty$ with and without linear DOB}

\begin{figure}
	\centering
	\includegraphics[width=\linewidth]{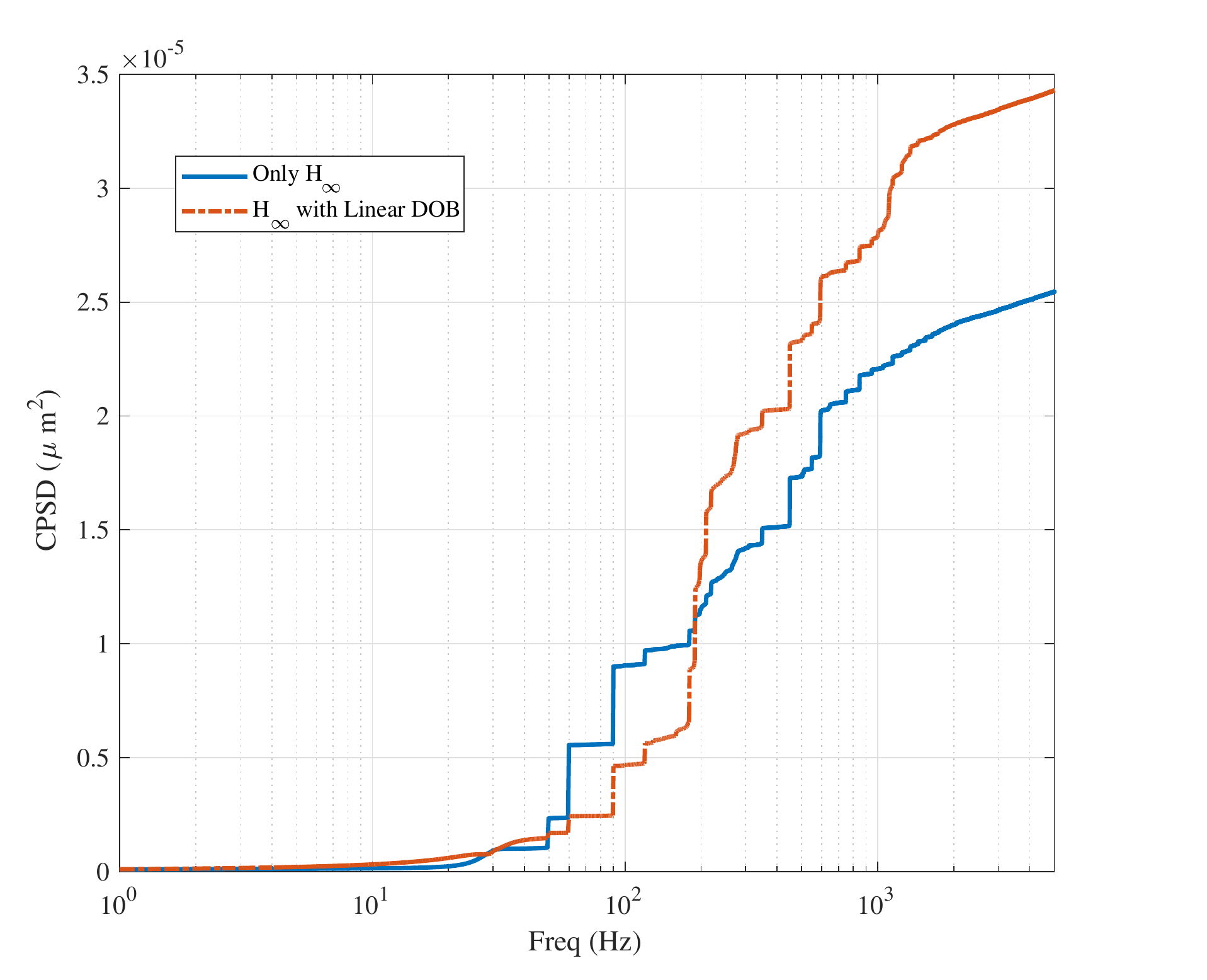}
	\caption{CPSD with $H_\infty$ controller with and without linear DOB in place}
	\label{fig:CPSD2}	
\end{figure}

CPSD of error signal in closed-loop with $H_\infty$ controller with and without linear DOB is provided in Fig. \ref{fig:CPSD2}. In all these experiments, noise is expected to be present at high frequencies ($500\ Hz$ or higher). No additional disturbance is applied to the setup. Instead, hysteresis of actuator acts as the disturbance in accordance with Fig. \ref{fig:DOB}. At low frequencies (below $228\ Hz$), use of linear DOB results in better performance and hysteresis compensation. However, at higher frequencies beyond the corner frequency of $Q$, the performance deteriorates quickly and is worse than the scenario where DOB is not used. In fact, it can be seen that the performance starts to deteriorate before corner frequency of $Q$ due to the inherent phase lag of linear filter.

\subsubsection{RDOB config-1}

\begin{figure}
	\centering
	\includegraphics[width=1\linewidth]{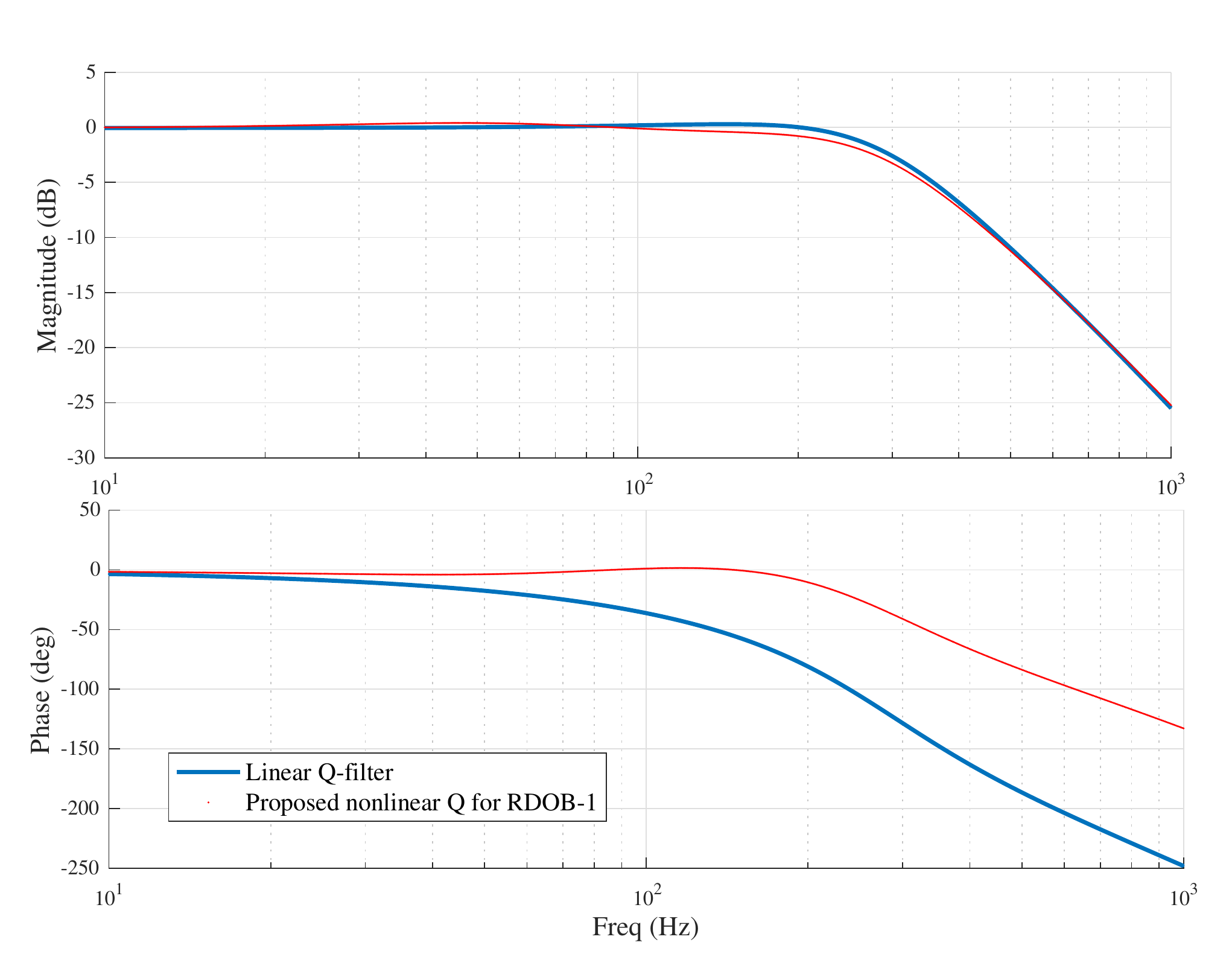}
	\caption{Frequency response of linear $Q$-filter compared with nonlinear $Q$-filter of RDOB-1 showing increase in phase close to corner frequency of $Q$}
	\label{fig:QfiltersDF}	
\end{figure}

\begin{figure}
	\centering
	\includegraphics[width=1\linewidth]{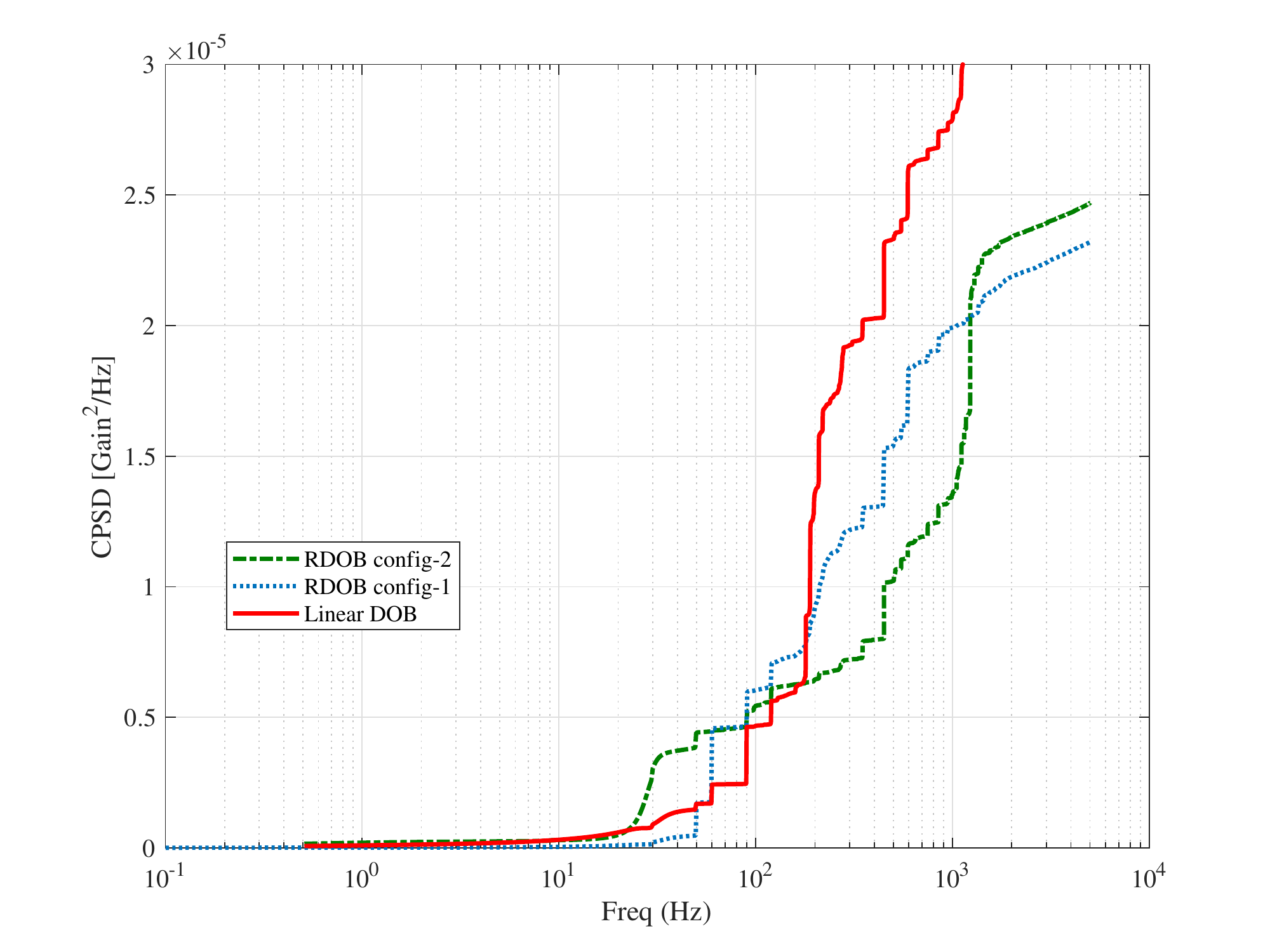}
	\caption{CPSD with proposed RDOB config-1 and RDOB config-2 along with that of linear DOB for comparison}
	\label{fig:set_1_CPSD}	
\end{figure}

\begin{figure}
	\centering
	\includegraphics[width=1\linewidth]{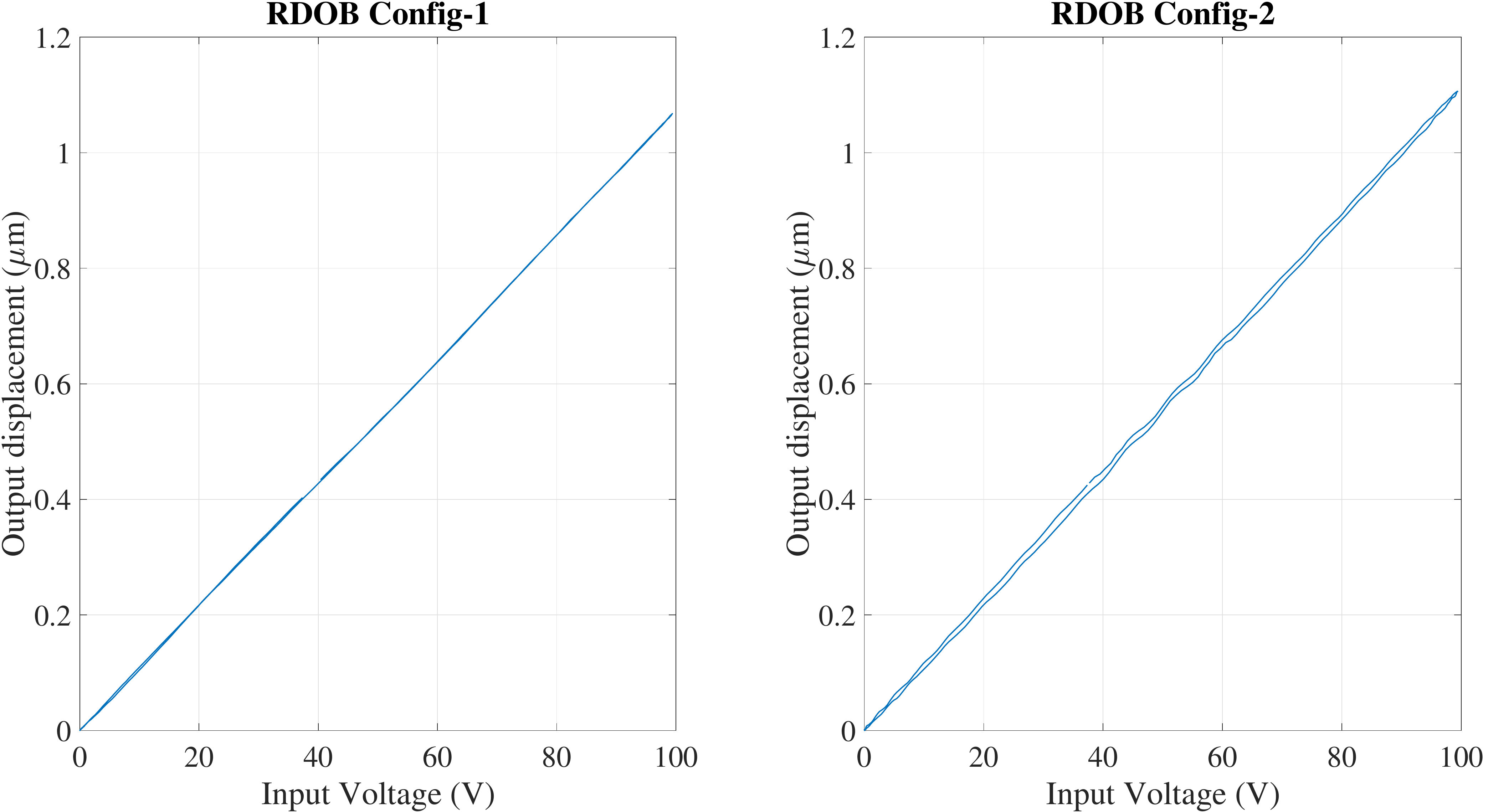}
	\caption{Hysteresis in the piezo-actuator at $30\ Hz$ for both RDOB configurations}
	\label{fig:hyst4a}	
\end{figure}

RDOB config-1 improves performance by adding phase in the inner loop and hence reducing phase lag introduced by $Q$-filter close to its corner frequency. This can be seen in Fig. \ref{fig:QfiltersDF}, where the frequency response of linear $Q$-filter as defined by Eqn. \ref{Qfilter} is compared with that of proposed nonlinear filter consisting of designed CgLp and as defined in Eqn. \ref{DOB1Qfilter}. The design of CgLp, $Q_1$ and $Q_2$ filters are as explained in Section. \ref{CgLpdesign}. CPSD of error for this configuration is shown in Fig. \ref{fig:set_1_CPSD}. It is seen that up to $200\ Hz$, performance of RDOB-1 is slightly worse or better than linear DOB depending on the frequency range considered, but the improvement due to added phase is clear beyond this where performance of RDOB-1 is significantly better than that of linear DOB. The hysteresis plot between input voltage and output displacement shown in Fig. \ref{fig:hyst4a} also indicates significant hysteresis compensation in RDOB-1.

\subsubsection{RDOB config-2}

\begin{figure}
	\centering
	\includegraphics[width=1\linewidth]{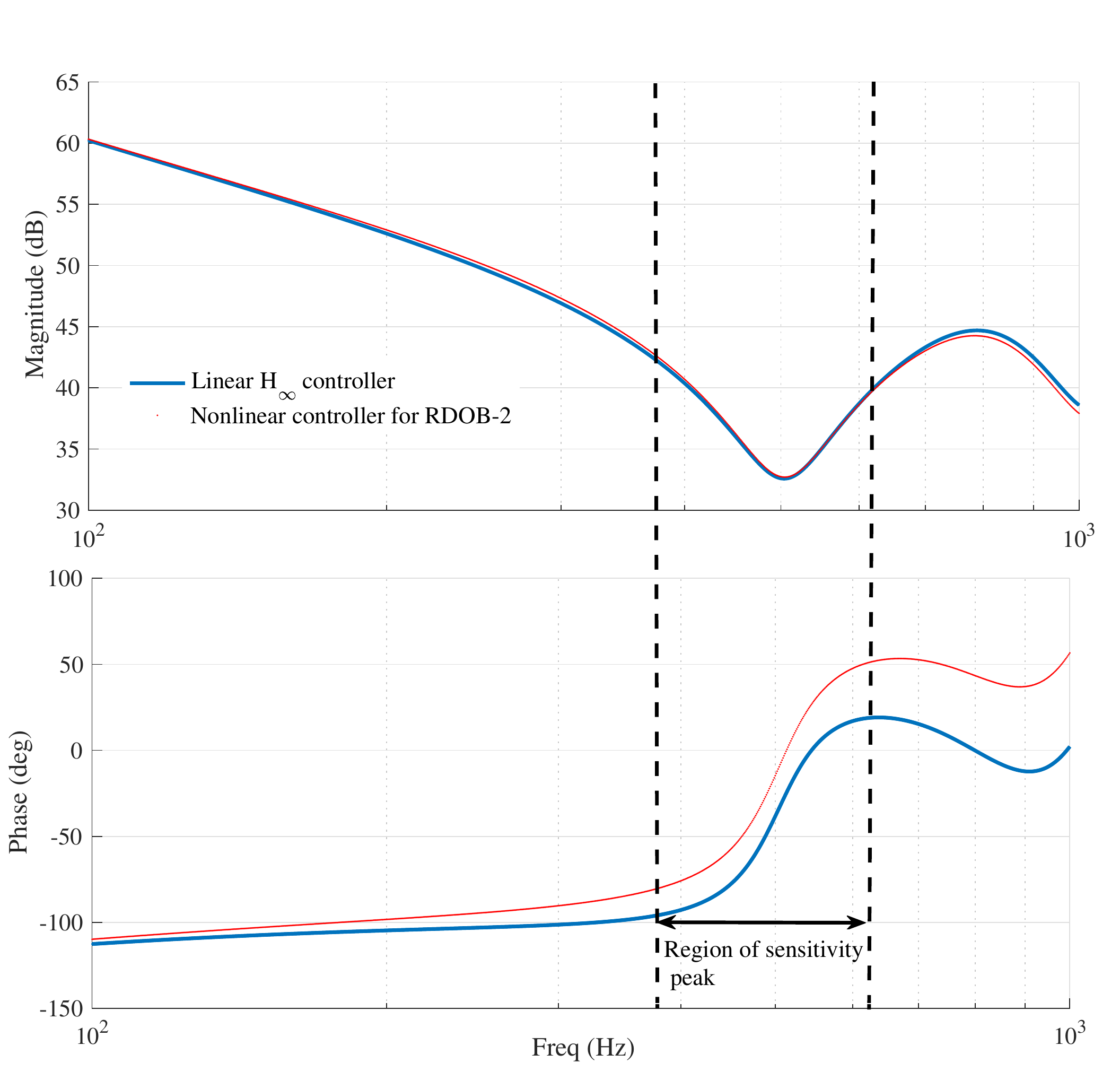}
	\caption{Frequency response of linear $H_\infty$ controller compared with nonlinear controller of RDOB-2 showing increase in phase in the region of sensitivity peak}
	\label{fig:controllerDF}	
\end{figure}

CgLp element is used in RDOB config-2 to improve phase of outer control loop close to the closed-loop bandwidth. The design of CgLp element used along with the $Q_{co}$ filter is as explained in Section. \ref{CgLpdesign}. This improvement in phase behaviour is shown in Fig. \ref{fig:controllerDF} with comparison between designed linear $H_\infty$ controller and proposed nonlinear controller consisting of $H_\infty$ with CgLp. CPSD for config-2 is also shown in Fig. \ref{fig:set_1_CPSD}. This shows slightly deteriorated performance at lower frequencies, but with greatly improved performance at higher frequencies where additional phase is provided by using CgLp element. This confirms the hypothesis that broadband phase compensation can lead to improved hysteresis compensation in the required band of frequencies. 

{
However, the hysteresis plot shown in Fig. \ref{fig:hyst4a} indicates that while noise rejection performance is better with RDOB-2, the same is not true of hysteresis compensation. This difference in hysteresis compensation is due to the difference in approach of the two configurations. In config-2, phase is added through CgLp to the controller loop to reduce peak of sensitivity. However this does not improve its disturbance rejection performance. However, in config-1, phase is added to inner loop to improve disturbance rejection and this improvement is clearly seen in the results. Since this difference in performance is due to differing design concepts, this variation can be expected at all applied reference frequencies.
}

\section{Conclusions and Future Work}

\subsection{Conclusions}

In this work, the fundamental limitations of linear disturbance observer are identified in an attempt to overcome the same using the nonlinear reset method. A novel CgLp element is introduced for broadband phase compensation and this is used in the two novel RDOB configurations presented. Following are the main conclusions drawn from this work :
\begin{itemize}
	\item The fundamental limitation of the linear DOB technique lies in linear theory and can only be overcome by introduction of nonlinearity.
	\item 
	CgLp element is used to overcome these limitations separately in two different configurations. An example plant and practical setup have been chosen to prove this.
	\item 
	
	{
	RDOB config-1 introduces CgLp in series with DEF $Q$ resulting in added phase and hence better disturbance rejection property as seen in results. Indirectly, this added phase also results in reduction of sensitivity peak.
}
	\item 
	
	{
	RDOB config-2 introduces CgLp in series with feedback controller resulting in reduction of sensitivity peak of outer loop and hence overall sensitivity peak.
}
	\item Both RDOB configurations perform better than linear DOB. However, while RDOB-1 works better at hysteresis compensation, RDOB-2 is better at noise attenuation.
	\item Appropriate configuration can be chosen based on application and requirements.
\end{itemize}

The concept of RDOB presented here is promising for improving performance of DOB. Adding phase through nonlinear CgLp element can be used to overcome fundamental limitations of linear DOB. However, more research is necessary to understand performance for optimal design of RDOB for required application. Both RDOB configurations are not optimally designed, but rather heuristically to validate the presented concepts.

\subsection{Recommendations}

Since resetting action results in the introduction of higher order harmonics, this might negatively affect system performance. This needs to be studied further to ensure good performance of RODB. Also, the trigger for resetting blocks is prone to noise and hence the level of noise can affect performance. Currently, a filter has been placed before the resetting CgLp element to avoid multiple resets possible due to noise. However, its effectiveness is not studied. Hence other techniques like fixed instant reset must be tested to avoid multiple resets without the use of additional filters.

\section{Acknowledgement}
Authors would like to thank ASML B.V. Netherlands for providing its facilities for this research. 
	
	\bibliographystyle{unsrtnat}
	\bibliography{ref2}

\end{document}